\documentstyle[amsfonts,epsf,10pt]{article}
\input{epsf}
\newcommand{\reell}{\kern+.23em\sf{I}\kern-.40em\sf{R}\kern+.76em\kern-.25em}

\def\n3{n_{3}}

\newcommand{\be}{\begin{equation}}
\newcommand{\ee}{\end{equation}}
\newcommand{\ba}{\begin{eqnarray}}
\newcommand{\ea}{\end{eqnarray}}
\input tcilatex
\begin{document}

\title{The Weyl representation on the torus}
\author{{\ A.M.F. Rivas\thanks{%
Corresponding author. e-mail: rivas@cbpf.br, Tel:(5521)5867223
fax:(5521)5867400 } and A.M. Ozorio de Almeida}}
\maketitle

\noindent

\centerline{\it Centro Brasileiro de Pesquisas F\'{\i}sicas} \centerline{\it %
Rua Xavier Sigaud 150, CEP 22290-180, RJ, Rio de Janeiro, Brazil}

\vspace{1cm} \noindent

\centerline {\bf Abstract} We construct reflection and translation operators
on the Hilbert space corresponding to the torus by projecting them from the
plane. These operators are shown to have the same group properties as their
analogue on the plane. The decomposition of operators in the basis of
reflections corresponds to the Weyl or center representation, conjugate to
the chord representation which is based on quantized translations. Thus, the
symbol of any operator on the torus is derived as the projection of the
symbol on the plane. The group properties allow us to derive the product law
for an arbitrary number of operators in a simple form. The analogy between the center and the chord
representations on the torus to those on the plane is then exploited to
treat Hamiltonian systems defined on the torus and to formulate a path
integral representation of the evolution operator. We derive its
semiclassical approximation.

\vspace{1cm}  \vfill

\clearpage

\section{Introduction}

The Weyl representation places quantum mechanics in phase space. The density
operator is mapped into a real phase function that projects onto the
position or momentum probability densities. The unitary operators
corresponding to linear canonical transformations transform the Weyl symbol
of any operator as a classical variable. Therefore, it is not surprising
that the study of the semiclassical limit of nonintegrable systems has
relied heavily on the Weyl representation as reviewed by reference \cite
{ozrep}.

The study of systems that are chaotic in the classical limit has developed
several models supported by a compact phase space. The simplest choice is
that of a $\left( 2L\right) $-dimensional torus, corresponding to a system
with $L$ degrees of freedom. Indeed, if the system propagates in discrete
time ( a mapping of the torus onto itself) even the special case $L=1$ may
be chaotic as is the case of the cat map \cite{arnold}, \cite{hanay} or the
baker's map \cite{balavoros}, \cite{saraceno}. It is well known that the
Hilbert space corresponding to a classical torus is finite. We may picture
the $N$ allowed positions and momenta as forming a lattice, the {\it quantum
phase space} (QPS), even though no rigorous definition of position and
momentum operators is available on the torus \cite{pqtor}. The semiclassical
limit is then obtained as $N\sim \left( 2\pi \hbar \right) ^{-1}\rightarrow
\infty .$

Though it is a great advantage to investigate numerically the propagation of
finite vectors or matrices defined in the QPS, it is somewhat disconcerting
that the difference between classical and quantum motion is more extreme on
the torus than on the plane. This problem also manifests itself in the
adaptation of the Weyl representation to the torus. The existing literature 
\cite{wooters}-\cite{kaperpeev} relies mainly on formal procedures, so that
the achievement of the classical limit as $N\rightarrow \infty $ may not be
preceded by the emergence of classical structures even for finite $\hbar $,
such as have been found for a plane phase space.

Clearly, a way to avoid this difficulty is to consider the classical torus
as a periodic plane phase space, to quantize the latter and then to project
this on QPS. We thus generalize the procedure of Hannay and Berry \cite
{hanay}, allowing for arbitrary ''Bloch '' or ''Floquet ''angles for each
circuit of the torus. It is then possible to project the appropriate plane
translation and reflection operators onto the torus. Hence, we define the
Weyl (or center) representation and its conjugate chord representation while
maintaining the main geometrical features characteristic of the plane.

In section 2 we present the translation operators of momentum and position
for a torus considered as the fundamental unit cell of the periodic plane.
By also allowing tori made up of more than one cell, we show that the unit
quantum torus may be obtained as the projection of the Hilbert space
corresponding to such a larger torus. Finally, by connecting the Hilbert
space for the plane to that of an infinitely large torus, we obtain the
projection of plane operators onto the torus.

Section 3 summarizes important features of the Weyl representation on the
plane. Defining the translation operators and their Fourier transform, the
reflection operators, we derive the Weyl symbols and their product rules in
terms of integrals over phase space polygons.

In section 4 we project appropriate translation operators onto QPS. However,
it is found that the reflection operators are supported on a lattice with
half the spacing of QPS . In consequence, the trace of these operators is
not homogeneous, which leads to complications in the formulae for products
of operators. Only in the case that $N$ is odd, can we simplify the product
rule for the Weyl symbols into a form that is analogous to the theory in the
plane. In any case, we need only half the number of sums derived in the
previous work of Galleti and Toledo Pisa \cite{galeti1} and these sums
depend on the symplectic areas of the same polygons arising in the plane
theory. Finally, we discuss the restricted form of symplectic invariance
that holds for the Weyl and chord representation on the torus: The Weyl
symbols transform classically under the action of quantum cat maps.

Though we have motivated our paper through discrete time models, periodic
Hamiltonians have obvious applications in solid state physics. Thus, section
5 is dedicated to the derivation of a path integral for the Weyl symbol of
the propagator. This relies on the symplectic area of polygons as in the
plane theory. In the semiclassical limit, the propagator is expressed in
terms of the center generating function presented in reference \cite{ozrep}.

Throughout this work we differentiate operators on the plane by italic $%
\widehat{A}$, as opposed to bold operators $\widehat{{\bf A}}$ on the torus.

\section{Hilbert space for tori}

\setcounter{equation}{0}

Classical phase space of ($2L)$ dimensions may be considered to be periodic,
so that we confer to it the topology of a $(2L)$ dimensional torus.
Evidently, we may use invariance, with respect to symplectic
transformations, to equate the periods $\Delta q=\Delta p=\nu $ of the
position and momentum coordinates. The usual choice is $\nu =1$, but we will
leave this as a free integer parameter so as to study the nesting of tori,
that is, the case where quantization is imposed on a larger (periodic)\
region than the unit cell. Thus, the number of unit cells will be $\nu
^{2L}. $

It is important to treat the specification of the Hilbert space of quantum
states for the torus, or {\it prequantization}, independently from the
dynamics of the system. That is, we treat the quantum kinematics,
corresponding to the geometrical description of phase space at the classical
level. A complete description for prequantization must include boundary
conditions; which are here that the wave functions satisfy Bloch conditions: 
\begin{eqnarray}
{\bf \Psi }(q+\nu ) &=&e^{2\pi i\chi _p}{\bf \Psi }(q),  \label{eq1} \\
{\bf \tilde{\Psi}}(p+\nu ) &=&e^{-2\pi i\chi _q}{\bf \tilde{\Psi}}(p)
\label{eq:eq2}
\end{eqnarray}
where 
\begin{equation}
{\bf \tilde{\Psi}}(p)=(2\pi \hbar )^{-1/2}\int e^{-ipq/\hbar }{\bf \Psi }%
(q)dq  \label{eq:91}
\end{equation}
and $2\pi i\chi _p$ and $2\pi i\chi _q$ are fixed arbitrary Floquet angles;
that is, the prequantization depends on the vector $\chi =(\chi _p,\chi _q)$
whose coordinates are in the range $0\le \chi _q,\chi _q<1$.

It is a well known kinematical restriction \cite{debievre} for torus
quantization that there are 
\begin{equation}
\nu ^2N=\nu ^2(2\pi \hbar )^{-1}  \label{eq:nh}
\end{equation}
basis states for each degree of freedom, so that $N=(2\pi \hbar )^{-1}$ is
the number of states corresponding to the unit cell .This is a crucial
point: the compactness of the phase space implies in the finiteness of the
dimension of the Hilbert space.

Recalling the translation operators $\widehat{T}_p=\exp \left( \frac{i\alpha 
}\hbar \widehat{q}\right) \,$ and $\widehat{T}_q=\exp \left( -\frac{i\beta }%
\hbar \widehat{p}\right) $ that respectively translates momentum by $\alpha $
and position by $\beta $ in the plane, we define minimal translators on the
torus $\widehat{{\bf T}}_{p,\nu ^2N}$ and $\widehat{{\bf T}}_{q,\nu ^2N}$
with their discrete eigenstates $|{\bf q}_n,\nu ^2N>$ and $|{\bf p}_m,\nu
^2N>$ such that 
\begin{equation}
\begin{array}{cc}
\widehat{{\bf T}}_{p,\nu ^2N}|{\bf q}_n,\nu ^2N>=e^{\left[ \frac{2\pi i}{\nu
^2N}(n+\chi _q)\right] }|{\bf q}_n,\nu ^2N> & \widehat{{\bf T}}_{p,\nu ^2N}|%
{\bf p}_m,\nu ^2N>=|{\bf p}_{m+1},\nu ^2N> \\ 
\widehat{{\bf T}}_{q,\nu ^2N}|{\bf q}_n,\nu ^2N>=|{\bf q}_{n+1},\nu ^2N> & 
\widehat{{\bf T}}_{q,\nu ^2N}|{\bf p}_m,\nu ^2N>=e^{\left[ -\frac{2\pi i}{%
\nu ^2N}(m+\chi _p)\right] }|{\bf p}_m,\nu ^2N>
\end{array}
\end{equation}

The product of $\nu ^2N$ translations on the basis states $|{\bf q}_n,\nu
^2N>$ or $|{\bf p}_m,\nu ^2N>\,$must return these to the same state, i.e.
these Schwinger operators \cite{schwinger} satisfy 
\begin{equation}
\left( \widehat{{\bf T}}_{p,\nu ^2N}\right) ^{\nu ^2N}=e^{2\pi i\chi _q}%
\widehat{{\bf 1}}_{\nu ^2N}^\chi 
\end{equation}
and 
\begin{equation}
\left( \widehat{{\bf T}}_{q,\nu ^2N}\right) ^{\nu ^2N}=e^{-2\pi i\chi _p}%
\widehat{{\bf 1}}_{\nu ^2N}^\chi \ .
\end{equation}

To define the Hilbert space ${\cal H}_{\nu ^2N}^\chi $, we add the Hermitian
structures 
\begin{equation}
<{\bf q}_n,\nu ^2N|{\bf q}_{n^{\prime }},\nu ^2N>=\delta _{n,n^{\prime
}}^{(\nu ^2N)}e^{\frac{2\pi i}{\nu ^2N}(n-n^{\prime })\chi _p},
\label{eq:QQsim}
\end{equation}
and 
\begin{equation}
<{\bf p}_m,\nu ^2N|{\bf p}_{m^{\prime }},\nu ^2N>=\delta _{m,m^{\prime
}}^{(\nu ^2N)}e^{-\frac{2\pi i}{\nu ^2N}(m-m^{\prime })\chi _q}.
\label{qbastgp}
\end{equation}
Here we define the {\it $N$-periodic Kronecker delta }

\begin{equation}
\delta _{m,n}^{(N)}\equiv \sum_{j=-\infty }^\infty \delta _{m,n+jN}.
\label{Nkrone}
\end{equation}
The bases are exchanged with the transformation kernel 
\begin{equation}
<{\bf p}_m,\nu ^2N|{\bf q}_n,\nu ^2N>=\frac 1{\nu N^{1/2}}e^{2\pi i\frac
1{\nu ^2N}(m+\chi _p)(n+\chi _q)}\equiv F_{m,n},  \label{eq:PQ}
\end{equation}
forming a unitary matrix ( finite Fourier transformation).

\begin{figure}[htb]
\centerline {\epsfxsize=5in  \epsffile{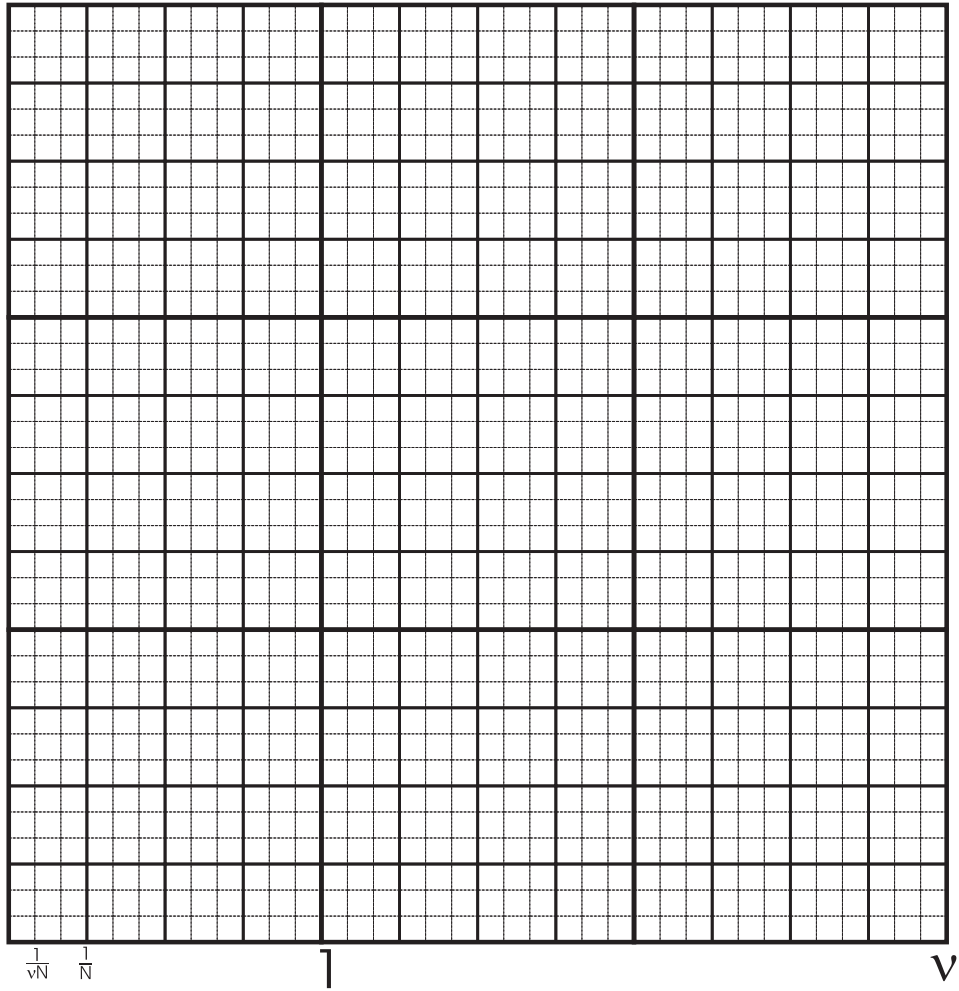} }
\caption{ The quantum Phase space for $N=4$. The intersection of the
bold lines in the unit square determine QPS for $\nu =1$, whereas the full
figure corresponds to the choice $\nu =3$.
}
\label{fig.3.1}
\end{figure}

Clearly, this last expression allows us to interpret the position ${\bf q}_n$
as corresponding to $q_n=\frac 1{\nu N}(n+\chi _q)$, whereas ${\bf p}_m$
corresponds to $p_m=\frac 1{\nu N}(m+\chi _p)$, leading to 
\begin{equation}
<{\bf p}_m,\nu ^2N|{\bf q}_n,\nu ^2N>=(2\pi \hbar )^{-\frac 12}\exp \left(
\frac i\hbar p_mq_n\right) .
\end{equation}
Likewise the $q$-Translator $\widehat{T}_q$, corresponds to a translation in
the plane by $\Delta q=\frac 1{\nu N}$ and the phase change $\exp \left[
2\pi i\chi _q\right] $ results from the translation $\Delta q=\nu $ $\,$%
around the torus. Although the indices $n$ and $m$ can run over all
integers, only $\nu ^2N$ successive values will form a basis for the torus
Hilbert space ${\cal H}_{\nu ^2N}^\chi .$ We may keep to the fundamental
range $\left[ 0,\nu ^2N-1\right] $, corresponding to the square with side $%
\nu $, or extend to the periodic plane, by taking into account the phases $%
\chi $. These considerations apply to each of the $L$ degrees of freedom, so
that in general the fundamental domain is a $\left( 2L\right) $-hypercube.
We see that position and momentum form a discrete web on the torus, as shown
in Fig.~\ref{fig.3.1}, that we call the quantum phase space (QPS), following reference 
\cite{galeti1}:

\begin{equation}
x=\left( 
\begin{array}{c}
p \\ 
q
\end{array}
\right) =\frac 1N\left( 
\begin{array}{c}
m+\chi _p \\ 
n+\chi _q
\end{array}
\right) .  \label{xtorop}
\end{equation}

We now consider the relation between the Hilbert spaces of two nested tori $%
{\cal H}_N^\chi $ and ${\cal H}_{\nu ^2N}^{\chi ^{\prime }}$ with $\nu >1.$
The consideration for the QPS corresponding to ${\cal H}_N^\chi $ to be a
sublattice of the QPS of ${\cal H}_{\nu ^2N}^{\chi ^{\prime }}$ is first
that 
\begin{equation}
\chi ^{\prime }=\nu \chi -k,  \label{nunup}
\end{equation}
where $k=(k_p,k_q)$ is an integer vector that denotes the number of loops
around the torus made by $\chi $ when multiplied by $\nu .$ Thus $\chi
^{\prime }$ is uniquely determined by $\chi .$ The indices for the larger
Hilbert space ${\cal H}_{\nu ^2N}^{\chi ^{\prime }}$ for points in the
fundamental domain of the smaller QPS are given by

\begin{eqnarray}
m^{\prime } &=&\nu m+k_p \\
n^{\prime } &=&\nu n+k_q.
\end{eqnarray}
We can now define the Hilbert space ${\cal H}_N^\chi $ as a projection of
the larger space ${\cal H}_{\nu ^2N}^{\chi ^{\prime }}.$ Indeed, it is easy
to verify that if $|{\bf q}_{n^{\prime }},\nu ^2N>$ are orthogonal position
eigenstates for ${\cal H}_{\nu ^2N}^{\chi ^{\prime }}$, then 
\begin{equation}
|{\bf q}_n,N>_\chi =\frac 1{\sqrt{\nu }}\sum_{r=0}^{\nu -1}e^{i2\pi \chi
_pr}|{\bf q}_{\nu n+k}+r,\nu ^2N>_{\chi ^{\prime }}  \label{eq:q1q2}
\end{equation}
form an appropriate orthonormal basis for ${\cal H}_N^\chi $. Thus, defining
the projection operator 
\begin{equation}
\widehat{{\bf 1}}_N^\chi =\sum_{n=0}^{N-1}|{\bf q}_n,N><{\bf q}_n,N|
\label{uno}
\end{equation}
we verify that this is a Hermitian operator in ${\cal H}_{\nu ^2N}^{\chi
^{\prime }}$, and that 
\begin{equation}
\widehat{{\bf 1}}_N^\chi \widehat{{\bf 1}}_N^\chi =\widehat{{\bf 1}}_N^\chi .
\end{equation}
Furthermore the states are obtained as 
\begin{equation}
|{\bf \Psi ,}N>=\widehat{{\bf 1}}_N^\chi {\bf |\Psi ,}\nu ^2N>.
\end{equation}

For all operators $\widehat{{\bf A}}_{\nu ^2N}$ acting in ${\cal H}_{\nu
^2N}^{\chi ^{\prime }},$ there is a projected operator which acts on ${\cal H%
}_N^\chi $: 
\begin{equation}
\widehat{{\bf A}}_N=\widehat{{\bf 1}}_N^\chi \widehat{{\bf A}}_{\nu ^2N}%
\widehat{{\bf 1}}_N^\chi .
\end{equation}
If $\widehat{{\bf A}}_{\nu ^2N}$ leaves ${\cal H}_N^\chi $ invariant, i.e.
if 
\begin{equation}
\lbrack \widehat{{\bf 1}}_N^\chi ,\widehat{{\bf A}}_{\nu ^2N}]=0,
\label{acomu}
\end{equation}
then 
\begin{equation}
\widehat{{\bf A}}_N=\widehat{{\bf A}}_{\nu ^2N}\widehat{{\bf 1}}_N^\chi .
\end{equation}
We also verify that, for any pair of operators satisfying (\ref{acomu}) 
\begin{equation}
\widehat{{\bf A}}_N\widehat{{\bf B}}_N=\widehat{{\bf A}}_{\nu ^2N}\widehat{%
{\bf B}}_{\nu ^2N}\widehat{{\bf 1}}_N^\chi .  \label{eq:abpt}
\end{equation}
In particular, we obtain the Schwinger translation operators for ${\cal H}%
_N^\chi $ as the projection of those in ${\cal H}_{\nu ^2N}^{\chi ^{\prime
}} $:

\begin{equation}
\widehat{{\bf T}}_{q,N}=\widehat{{\bf T}}_{q,\nu ^2N}\widehat{{\bf 1}}%
_N^\chi \qquad ;\qquad \widehat{{\bf T}}_{p,N}=\widehat{{\bf T}}_{p,\nu ^2N}%
\widehat{{\bf 1}}_N^\chi .
\end{equation}

Let us now take the limit $\nu \rightarrow \infty $. Clearly, the variable $%
q_n$ becomes continuous in this limit as the volume of the torus $\nu
^{2L}\rightarrow \infty $.Throughout the limit, the relation (\ref{nunup})
defines an appropriate $\chi ^{\prime }\left( \chi \right) $. The main step
to recover the Banach space ${\cal H}_{{\Bbb R}}$ for the plane $\left(
2L\right) $-dimensional phase space is to redefine the normalization so that 
\begin{equation}
<q|q^{\prime }>=\delta (q-q^{\prime }),  \label{qbastg2}
\end{equation}
introducing the Dirac delta function on the right, and the continuous
Fourier integral for the change of basis: 
\begin{equation}
<p|\psi >=\left( 2\pi \hbar \right) ^{-L/2}\int dq\exp \left( \frac{ipq}%
\hbar \right) <q|\psi >
\end{equation}
In all other respects, the kinematics in the plane will coincide with that
of an infinitely large torus. However, the normalization condition (\ref
{qbastg2}) implies a change in the way we express the sates in ${\cal H}%
_N^\chi $ in terms of those of ${\cal H}_{{\Bbb R}}$. For that purpose, we
recall that for an unit torus 
\begin{equation}
\sum_{r=0}^{N-1}e^{i\frac{2\pi }N(m-n)r}=N\delta _{m,n}^{(N)},
\label{eq:114}
\end{equation}
so we extend the definition of the $N$-periodic Kroeneker delta function to
real numbers $x$ and $y$: 
\begin{equation}
\delta _{x,y}^{(N)}\equiv \left\langle e^{i\frac{2\pi }N\left( x-y\right)
k}\right\rangle _k,  \label{eq:kroneinf}
\end{equation}
where $\left\langle ...\right\rangle _k$ denotes the average over $k,$ 
\begin{equation}
\left\langle ...\right\rangle _k=\lim_{r\rightarrow \infty }\frac
1r\sum_{k=-\frac r2}^{\frac r2}.  \label{kprom}
\end{equation}
From (\ref{eq:kroneinf}) we see that $\delta _{x,y}^{(N)}$ only depends on
the difference $x-y$, so let us take $y=0$ for simplicity. For $x=r$ an
integer number, the argument in (\ref{eq:kroneinf})\ has period $N$. Then,
the average is just (\ref{eq:114}) divided by $N,$ so the definition (\ref
{eq:kroneinf})\ is consistent with (\ref{Nkrone}) for integer arguments. Let
us now suppose $x=\frac rd,$ a rational number. The argument in (\ref
{eq:kroneinf})\ thus has period $Nd$, so that 
\begin{equation}
\delta _{x,0}^{(N)}=\frac 1{Nd}\sum_{k=0}^{Nd-1}e^{i\frac{2\pi }N\frac
rdk}=\frac 1{Nd}\frac{1-e^{i2\pi r}}{1-e^{i\frac{2\pi }{Nd}r}}=\left\{ 
\begin{array}{cc}
1 & \mbox{ if }r=0\qquad \mbox{mod}(Nd) \\ 
0 & \mbox{ otherwise}
\end{array}
\right. .  \label{krorac}
\end{equation}
Hence, once more the $N$-periodic Kroeneker delta function is different from
zero only for $x$ being zero modulo $N$. By allowing $d\rightarrow \infty $
in (\ref{krorac}) we can extend the definition of $\delta _{x,y}^{(N)}$ to
irrational numbers, so that 
\begin{equation}
\delta _{x,y}^{(N)}=\left\{ 
\begin{array}{cc}
1 & \mbox{ only if }\left( x-y\right) =0\qquad \mbox{mod}(N) \\ 
0 & \mbox{ otherwise}
\end{array}
\right. .
\end{equation}

The definition (\ref{eq:kroneinf}) is an interpolation of (\ref{Nkrone}),
which will allows us to perform, not only sums with the $N$ periodic
Kroeneker delta function, but also integrals. Indeed, we will relie on the
formal equivalence: 
\begin{equation}
\delta _{x,y}^{(N)}=\left\langle \delta (\frac{x-y}N-k)\right\rangle _k.
\label{eq:dirkro}
\end{equation}
This is a consequence of the definition(\ref{eq:kroneinf}) and the Poisson
sum formula, 
\begin{equation}
\sum_{t\in {\Bbb Z}}\delta (x-t)=\sum_{k\in {\Bbb Z}}e^{i2\pi kx}.
\label{eq:poisson}
\end{equation}
Indeed, from the definition (\ref{eq:kroneinf}) we have 
\begin{equation}
\frac 1\nu \sum_{k=-\frac \nu 2}^{\frac \nu 2}e^{i\frac{2\pi }N\left(
x-y\right) k}=\frac 1\nu \sum_{k=-\frac \nu 2}^{\frac \nu 2}\delta (\frac{x-y%
}N-k)+\frac 1\nu R_\nu (x-y)
\end{equation}
with 
\begin{equation}
R_\nu (x-y)=\sum_{k=-\frac \nu 2}^{\frac \nu 2}e^{i\frac{2\pi }N\left(
x-y\right) k}-\sum_{k=-\frac \nu 2}^{\frac \nu 2}\delta (\frac{x-y}N-k).
\end{equation}
From the Poisson sum formula (\ref{eq:poisson}), we have that, $\lim_{\nu
\rightarrow \infty }R_\nu (x-y)=0,$ upon an appropriate ordering of the
limits $\nu \rightarrow \infty $ and the width of the delta function$%
\rightarrow 0$.A\ consequence of (\ref{eq:dirkro}) is that, for any function 
$f(t),$ 
\begin{equation}
\left\langle \delta (\frac{x-y}N-t)f(t)\right\rangle _t=\delta
_{x,y}^{(N)}\;f(\frac{x-y}N).  \label{eq:dirkrox}
\end{equation}

Changing the origin, so as to keep the unit torus at the center of the
larger torus, (\ref{eq:q1q2}) must be replaced by 
\begin{equation}
|{\bf q}_n,N>=\lim_{\nu \rightarrow \infty }\frac 1\nu \sum_{r=-\frac \nu
2}^{\frac \nu 2}|\frac{n+\chi _q}N+r>e^{2\pi ir\chi _p}.  \label{qpltor}
\end{equation}
A straightforward calculation using (\ref{eq:dirkrox}) shows that the
orthonormality conditions (\ref{eq:QQsim}) are obtained for the states
defined in (\ref{qpltor}) with the normalization (\ref{qbastg2}). So we
consider that the states and the operators in the unit torus are obtained
from the plane by projections. Recalling the definition (\ref{kprom}) of the
average, (\ref{qpltor}) can be written as 
\begin{equation}
|{\bf q}_n,N>=\left\langle |\frac{n+\chi _q}N+r>e^{2\pi ir\chi
_p}\right\rangle _r  \label{eq:qQ}
\end{equation}
and the projection operator $\widehat{{\bf 1}}_N^\chi \,$ is then given by (%
\ref{uno}). In the case of $\chi =0\,$ we thus retrieve the definition of
Hannay and Berry \cite{hanay} for ${\cal H}_N^0$ as the average over a
periodic array of Dirac delta distributions. We will now derive the Weyl
representation on the torus by projecting the properties that have been well
established in the plane.

\section{The Weyl representation on the plane}

\setcounter{equation}{0}

We here summarize the results obtained for the plane in reference \cite
{ozrep} that will be projected onto the torus in the following sections. We
define the operator corresponding to a general translation in phase space by
the $\left( 2L\right) $-dimensional vector $\xi =(\xi _p,\xi _q)$ as 
\begin{equation}
\hat{T}_\xi \equiv \exp \left( \frac i\hbar \xi \wedge \hat{x}\right) \equiv
\exp \left[ \frac i\hbar (\xi _p.\hat{q}-\xi _q.\hat{p})\right] ,
\label{eq:tcor}
\end{equation}
where naturally $\hat{x}=(\hat{p},\hat{q})$ and the symplectic product $\xi
\wedge \eta $ is defined as 
\begin{equation}
\xi \wedge \eta =({\frak J}\xi ).\eta
\end{equation}
with 
\begin{equation}
{\frak J}=\left[ 
\begin{array}{c|c}
0 & -1 \\ \hline
1 & 0
\end{array}
\right] .
\end{equation}
$\hat{T}_\xi $ is also known as a {\it Heisenberg operator}. In the case
where either $\xi _p$ or $\xi _q=0$, we obtain respectively the operators $%
\hat{T}_q$ or $\hat{T}_p$ mentioned in the preceding section.

Acting on the Banach space ${\cal {H}_{{\Bbb R}}}$ we have 
\begin{equation}
\widehat{T}_\xi |q_a>=e^{\frac i\hbar \xi _p(q_a+\frac{\xi _q}2)}|q_a+\xi _q>
\label{eq:tq}
\end{equation}
and

\begin{equation}
\widehat{T}_\xi |p_a>=e^{-\frac i\hbar \xi _q(p_a+\frac{\xi _p}2)}|p_a+\xi
_p>.
\end{equation}

The classical group property is maintained within a phase factor: 
\begin{equation}
\hat{T}_{\xi _2}\hat{T}_{\xi _1}=\hat{T}_{\xi _1+\xi _2}\ \exp [\frac{-i}{%
2\hbar }\xi _1\wedge \xi _2]=\hat{T}_{\xi _1+\xi _2}\ \exp [\frac{-i}\hbar
D_3(\xi _1,\xi _2)],  \label{eq:tt}
\end{equation}
where $D_3$ is the symplectic area of the triangle determined by two of its
sides. Evidently, the inverse of the unitary operator $\hat{T}_\xi ^{-1}=%
\hat{T}_\xi ^{\dag }=\hat{T}_{-\xi }$ and we can generalize (\ref{eq:tt}):

\begin{equation}
\widehat{T}_{\xi _1}...\widehat{T}_{\xi j}=\widehat{T}_{\xi _1+....+\xi
_j}e^{\frac i\hbar D_{j+1}(\xi _1,....\xi _j)},  \label{eq:ttt}
\end{equation}
where $D_{j+1}(\xi _1,....\xi _j)$ denotes the symplectic area of the $%
\left( j+1\right) $ sided polygon formed by the chords, $(\xi _1,....\xi _j)$%
.

The operator corresponding to a phase space reflection about a point $%
x=(p,q) $ is \cite{ozrep} 
\begin{equation}
\widehat{R}_x\equiv (4\pi \hbar )^{-L}\int d\xi \quad e^{\frac i\hbar
x\wedge \xi }\widehat{T}_\xi .  \label{eq:rint}
\end{equation}
This operator has the following properties \cite{ozrep} : 
\begin{equation}
\widehat{T}_\xi =(\pi \hbar )^{-L}\int dx\quad e^{-\frac i\hbar x\wedge \xi }%
\widehat{R}_x\ ,  \label{eq:tintr}
\end{equation}
\begin{equation}
\widehat{R}_x\widehat{T}_\xi =\widehat{R}_{x-\xi /2}e^{-\frac i\hbar x\wedge
\xi }\ ,  \label{eq:rt}
\end{equation}
\begin{equation}
\widehat{T}_\xi \widehat{R}_x=\widehat{R}_{x+\xi /2}e^{-\frac i\hbar x\wedge
\xi }\   \label{eq:tr}
\end{equation}
and 
\begin{equation}
\widehat{R}_{x_1}\widehat{R}_{x_2}=\widehat{T}_{2(x_2-x_1)}e^{\frac i\hbar
2x_1\wedge x_2},  \label{eq:rr}
\end{equation}
so that 
\begin{equation}
\widehat{R}_x\widehat{R}_x=\widehat{1}\ .
\end{equation}

The trace of the translation is 
\begin{eqnarray}
Tr\hat{T}_\xi &=&\int <q|T_\xi |q>dq=(2\pi \hbar )^L\delta (\xi _p)\delta
(\xi _q)  \nonumber \\
&=&(2\pi \hbar )^L\delta (\xi )\ 
\end{eqnarray}
and then taking the Fourier transform, 
\begin{equation}
Tr\hat{I\!\!\!\!R}_x=Tr2^L\hat{R}_x=(2\pi \hbar )^{-L}\int d\xi \exp \left[
\frac i\hbar \ x\wedge \xi \right] Tr\hat{T}_\xi =1\ ,
\end{equation}
where it is now also convenient to define the exact Fourier transform $\hat{%
I\!\!\!\!R}_x$ of $\hat{T}_\xi $. We recall that the classical
transformation $R_x$ has a single fixed point ($x$ itself), whereas $T_\xi $
has fixed points only if $\xi =0$, when all points are fixed. These results
are in general agreement with our intuition as to the classical
correspondence of the traces of unitary operators. i.e. that the trace is
related to the classical fixed points.

The above properties allow any operator $\hat{A}$ to be expressed as a
linear superposition of elementary translation operators: 
\begin{equation}
\hat{A}=\int \frac{d\xi }{(2\pi \hbar )^L}\ A(\xi )\hat{T}_\xi \ .
\label{eq:cora}
\end{equation}
The confirmation results from 
\begin{equation}
Tr(\hat{T}_{-\xi }\hat{A})=A(\xi )\ ,  \label{eq:acor}
\end{equation}
The analogy with the classical chord generating function for canonical
transformations is discussed in \cite{ozrep}. We can equally represent any
operator $\hat{A}$ as a superposition of reflections: 
\begin{equation}
\hat{A}=\int \frac{dx}{(2\pi \hbar )^L}\ A(x)\hat{I\!\!\!\!R}_x=\int \frac{dx%
}{(\pi \hbar )^L}\ A(x)\hat{R}_x\ .  \label{eq:cena}
\end{equation}
Again we obtain the expansion coefficient by calculating 
\begin{equation}
Tr(\hat{I\!\!\!\!R}_x\hat{A})=A(x)\ .  \label{eq:acen}
\end{equation}
Notice that comparison of (\ref{eq:cora}) and (\ref{eq:cena}) with (\ref
{eq:rint}) and (\ref{eq:tintr}) yields 
\begin{equation}
R_x(\xi )=2^{-L}\exp \left[ \frac i\hbar \ x\wedge \xi \right] \ \ \ %
\mbox{and}\ \ \ T_\xi (x)=\exp \left[ -\frac i\hbar \ x\wedge \xi \right] \ .
\end{equation}
In analogy with our previous result, we may refer to $A(x)$ as the {\it %
center representation} of the operator $\hat{A}$, but the historic term is
the {\it Weyl representation}.

The product laws of center and chord representations are of fundamental
interest. Starting with the chord representation, we have, for the product $%
\hat{A}_n\hat{A}_{n-1}\cdots \hat{A}_1$, 
\begin{eqnarray}
&&A_n.A_{n-1}\cdots A_1(\xi )=\left( \frac 1{2\pi \hbar }\right)
^{L(n-1)}\int d\xi _n\cdots d\xi _1A_n(\xi _n)\cdots  \nonumber \\
&&A_1(\xi _1)\delta (\xi _1+\cdots \xi _n-\xi )\exp \left[ -\frac i\hbar
D_{n+1}(\xi _1,\cdots ,\xi _n)\right] ,  \label{eq:ancor}
\end{eqnarray}
where we note that the Dirac $\delta $-function has reduced the $(n+2)$%
-sided polygon with symplectic area $D_{n+2}$ to an $(n+1)$-sided polygon,
with $n$ free sides. Evidently, we can now use the $\delta $-function to
remove one of the variables in the integral, but (\ref{eq:ancor}) is in its
most symmetric form.


We shall also need integral formulae for the product of operators in the
center representation. The result depends crucially on the parity of the
number of operators\cite{ozrep}, so we will start with the simplest case
where $n=2$. Proceeding from the definition ~(\ref{eq:cena}), we obtain 
\begin{equation}
A_2.A_1(x)=\left( \frac 1{\pi \hbar }\right) ^{2L}\int
dx_2dx_1A_2(x_2)A_1(x_1)\exp \left[ \frac i\hbar \Delta _3(x,x_1,x_2)\right]
.  \label{eq:aacen}
\end{equation}
where $\Delta _3(x,x_1,x_2)=2(x_1\wedge x_2+x_2\wedge x+x\wedge x_1)$ is the
symplectic area of the triangle whose midpoints are, $x,x_1,x_2$.

The extension to $(2n)$ operators is \cite{ozrep} 
\begin{eqnarray}
&&A_{2n}\cdots A_1(x)=  \nonumber \\
&&\left( \frac 1{\pi \hbar }\right) ^{2nL}\int dx_{2n}\cdots
dx_1A_{2n}(x_{2n})\cdots A_1(x_1)\exp \left\{ \frac i\hbar \Delta
_{2n+1}(x,x_1,\cdots ,x_{2n})\right\} .  \label{eq:ancen}
\end{eqnarray}
Here the symplectic area $\Delta _{2n+1}$ corresponds to the $(2n+1)$-sided
polygon circumscribed around the centers $(x,x_1,\cdots ,x_{2n})$. 

The main advantage of the chord and center representation is their
symplectic invariance. It is well known that linear classical canonical
transformations $x^{\prime }=Mx$ correspond to unitary transformations in $%
{\cal {H}_{{\Bbb R}}}$%
\begin{equation}
\widehat{A}\rightarrow \widehat{U}_M\widehat{A}\widehat{U}_M^{-1}.
\end{equation}
The effect of such a unitary transformation on the chord and center
representation is merely 
\begin{equation}
A(x)\rightarrow A(Mx)\quad \mbox{and}\quad A(\xi )\rightarrow A(M\xi ).
\end{equation}

\section{Weyl representation in the torus}

\setcounter{equation}{0}

\subsection{ Translation and Reflection Operators on the torus}

In this section we will project the translations $\widehat{T}_\xi $ and
reflections $\widehat{R}_x$ operators defined on ${\cal H}_{{\Bbb R}}$ onto
the Hilbert space ${\cal H}_N^\chi $. Again we will treat the case of one
degree of freedom explicitly, since the generalization is obvious. For this
purpose we first investigate the action of the translation operators $%
\widehat{T}_\xi $ on the $|{\bf q}_n,N>$ basis vectors defined by (\ref
{eq:qQ}). From the effect of a translation (\ref{eq:tq}) on a single
position in the plane, we have 
\begin{equation}
\widehat{T}_\xi |{\bf q}_n,N>=\left\langle |\frac{n+\chi _q}N+k+\xi
_q>e^{2\pi ik\chi _p}e^{\frac i\hbar \xi _p(\frac{n+\chi _q}N+k+\frac 12\xi
_q)}\right\rangle _k,
\end{equation}
using the relation (\ref{eq:nh}) between $N$ and $\hbar $. This vector will
belong to ${\cal H}_N^\chi $ only if it has the form (\ref{eq:qQ}), i.e., if
we can write $\xi _q=\frac sN$ with $s$ an integer.

A similar treatment in the $|{\bf p}_m,N>$ representation implies that 
\begin{equation}
\widehat{T}_\xi |{\bf p}_m,N>=\left\langle |\frac{m+\chi _p}N+k+\xi
_p>e^{-2\pi ik\chi _q}e^{-\frac i\hbar \xi _q(\frac{m+\chi _q}N+k+\frac{\xi
_p}2)}\right\rangle _k,
\end{equation}
which does not belong to ${\cal H}_N^\chi $ unless we can write $\xi
_p=\frac rN$ with $r$ an integer. So, as was already pointed in \cite
{debievre}, the only translations that leave ${\cal H}_N^\chi $ invariant
are those whose chords are $\xi =(\frac rN,\frac sN)$. For these cases we
have 
\begin{eqnarray}
\widehat{T}_\xi |{\bf q}_n,N &>&=\left\langle |\frac{n+\chi _q}N+k+\frac
sN>e^{2\pi ik\chi _p}e^{i2\pi N\frac rN(\frac{n+\chi _q}N+k+\frac
s{2N})}\right\rangle _k  \nonumber \\
&=&e^{\frac{i2\pi }Nr(n+\frac s2+\chi _q)}\left\langle |\frac{n+s+\chi _q}%
N+k>e^{2\pi ik(\chi _p+r)}\right\rangle _k  \nonumber \\
&=&e^{\frac{i2\pi }Nr(n+\frac s2+\chi _q)}|{\bf q}_{n+s},N>.  \label{eq:tQ}
\end{eqnarray}

In short, we obtain the torus operator $\widehat{{\bf T}}_\xi ^\chi $ in
terms of the plane operator $\widehat{T}_\xi $ as 
\begin{eqnarray}
\widehat{{\bf 1}}_N^\chi \widehat{T}_\xi \widehat{{\bf 1}}_N^\chi &=&\left\{ 
\begin{tabular}{lll}
$\widehat{{\bf T}}_\xi ^\chi $ & $\mbox{if }$ $\xi =\left( \frac rN,\frac
sN\right) ,$where $r$ and $s$ are integers &  \\ 
$0$ & otherwise & 
\end{tabular}
\right. \\
&=&\widehat{{\bf T}}_\xi ^\chi \left\langle \delta \left( \xi -\frac
1N\left( r,s\right) \right) \right\rangle _{\left( r,s\right) }=\widehat{T}%
_\xi \widehat{{\bf 1}}_N^\chi \ ,  \label{projt}
\end{eqnarray}
where the torus translation operators $\widehat{{\bf T}}_\xi ^\chi \equiv 
\widehat{{\bf T}}_{r,s}^\chi $ are defined through 
\begin{equation}
\widehat{{\bf T}}_{r,s}^\chi |{\bf q}_n,N>=e^{i\frac{2\pi }Nr(n+\chi
_q+s/2)}|{\bf q}_{n+s},N>  \label{eq:TQ}
\end{equation}
and 
\begin{equation}
\widehat{{\bf T}}_{r,s}^\chi |{\bf p}_m,N>=e^{-i\frac{2\pi }Ns(m+\chi
_p+r/2)}|{\bf p}_{m+r},N>.  \label{eq:TP}
\end{equation}
The last equality in (\ref{projt}) holds because $\widehat{T}_\xi $ and $%
\widehat{{\bf 1}}_N^\chi $ commute for $\xi =\left( \frac rN,\frac sN\right) 
$. We then see that the only translation operators that do not vanish on
projection to the torus are those that leave ${\cal H}_N^\chi $ invariant.
They correspond precisely to those classical transformations that preserve
the QPS web. To simplify the notation, we will usually assume implicitly the 
$\chi $ dependence.

Let us now study some properties of the torus translation operators. For the
case where the chords are the minimal translations in any one of the $q$ or $%
p$ directions, we recover the Schwinger operators \cite{schwinger} so that,

\begin{equation}
\widehat{{\bf T}}_{0,1}|{\bf q}_n,N>\equiv \widehat{{\bf T}}_q|{\bf q}_n,N>=|%
{\bf q}_{n+1},N>
\end{equation}
and, 
\begin{equation}
\widehat{{\bf T}}_{1,0}|{\bf p}_m,N>\equiv \widehat{{\bf T}}_p|{\bf p}_m,N>=|%
{\bf p}_{m+1},N>.
\end{equation}
The kernel (\ref{eq:PQ}) implies that, 
\begin{equation}
\widehat{{\bf T}}_p\widehat{{\bf T}}_q=e^{-\frac{2\pi i}N}\widehat{{\bf T}}_q%
\widehat{{\bf T}}_p.  \label{eq:grupo}
\end{equation}
so any translation operator in ${\cal H}_N^\chi $ is defined as 
\begin{equation}
\widehat{{\bf T}}_\xi \equiv \widehat{{\bf T}}_{r,s}=e^{-\frac{i\pi rs}N}%
\widehat{{\bf T}}_p^r\widehat{{\bf T}}_q^s\ ,  \label{eq:Tcor}
\end{equation}
with chords $\xi =(\frac rN,\frac sN).$ We can express the matrix elements
of the translation operators in the $|{\bf q}_n,N>$ basis,

\begin{equation}
<{\bf q}_m,N|\widehat{{\bf T}}_{r,s}|{\bf q}_n,N>=e^{-i\frac{2\pi }Nr(\frac{%
m+n}2+\chi _q)}\delta _{m,n+s}^{(N)}e^{i\frac{2\pi }N(\frac r2+\chi
_p)(m-n-s)},  \label{eq:TQQ}
\end{equation}
using the orthonormality relations of the states (\ref{eq:QQsim}).The fact
that the Hilbert space has finite dimension implies that linear operators
acting on it will be represented by $N\times N$ matrices. Then $N^2$
linearly independent matrices will form a basis for the operators in ${\cal H%
}_N^\chi $. This is clear from the symmetries of the translation operators;
through their action on ${\cal H}_N^\chi $ (\ref{eq:TQ}) we see that

\begin{equation}
\widehat{{\bf T}}_{\xi +{\bf k}}=(-1)^{sk_p-rk_q+k_pk_qN}e^{i2\pi (k_p\chi
_q-k_q\chi _p)}\widehat{{\bf T}}_\xi =e^{-i2\pi N\left[ \left( \frac \xi
2+\frac \chi N\right) \wedge {\bf k}+\frac 14{\bf k}\widetilde{{\frak J}}%
{\bf k}\right] }\widehat{{\bf T}}_\xi ,  \label{eq:TTsim}
\end{equation}
where ${\bf k}=(k_p,k_q)$ is a vector with integer components denoting
chords that perform respectively $k_p$ and $k_q$ loops around the
irreducible circuits of the torus. We have also defined the symmetric matrix 
\begin{equation}
\widetilde{{\frak J}}=\left[ 
\begin{array}{c|c}
0 & 1 \\ \hline
1 & 0
\end{array}
\right] .
\end{equation}
If we perform ${\bf k}$ loops around the torus, (\ref{eq:TTsim}) implies
that we recover the identity operator only up to a phase: 
\begin{equation}
\widehat{{\bf T}}_{{\bf k}}=(-1)^{k_pk_qN}e^{i2\pi (k_p\chi _q-k_q\chi _p)}%
\widehat{{\bf 1}}_N^\chi .  \label{tNQ}
\end{equation}
Thus, to have a basis of operators we only need $r$ and $s$ in the range $%
[0,N-1]$, that is, we only need translations that perform less than one loop
around the torus.

The second phase factor in expression (\ref{eq:TTsim}) comes from the Bloch
boundary conditions, but the $(-1)^{sk_p-rk_q+k_pk_qN}$ factor shows that we
need two loops around the torus to recover the same operator, doubling the
expected periodicity. This will have crucial importance in the construction
of the reflection operators.

An important property of the translation operators, which can be deduced
from (\ref{eq:grupo}) is 
\begin{equation}
\widehat{{\bf T}}_{\xi _2}\widehat{{\bf T}}_{\xi _1}=\widehat{{\bf T}}_{\xi
_1+\xi _2}e^{i\pi N\xi _1\wedge \xi _2},  \label{eq:TT}
\end{equation}
which generalizes to 
\begin{equation}
\widehat{{\bf T}}_{\xi _1}...\widehat{{\bf T}}_{\xi j}=\widehat{{\bf T}}%
_{\xi _1+....+\xi _j}e^{i\pi ND_{j+1}(\xi _1,....\xi _j)},
\end{equation}
where $D_{j+1}(\xi _1,....\xi _j)$ denotes the area of $j+1$ sided polygon
formed by the chords, $(\xi _1,....\xi _j)$, exactly as (\ref{eq:ttt}) in
the plane case. From (\ref{eq:TT}) and the unitarity of $\widehat{{\bf T}}%
_\xi ,$ we can see that 
\begin{equation}
\widehat{{\bf T}}_\xi ^{\dag }=\widehat{{\bf T}}_\xi ^{-1}=\widehat{{\bf T}}%
_{-\xi }.  \label{eq:T-1}
\end{equation}
Notice that  (\ref{eq:TT}) reduces to (\ref{eq:grupo}) for the particular case that $\xi _1$ and $\xi _2$ are vectors along the coordinate axis.

For the reflection operators, we use their definition (\ref{eq:rint}) in
terms of plane translations and then (\ref{projt}) projects the translations
onto the torus, so that 
\begin{eqnarray}
\widehat{{\bf R}}_x^\chi &=&\widehat{{\bf 1}}_N^\chi \widehat{R}_x\widehat{%
{\bf 1}}_N^\chi =(4\pi \hbar )^{-L}\int d\xi \quad e^{\frac i\hbar x\wedge
\xi }\widehat{{\bf T}}_\xi \left\langle \delta \left( \xi -\frac{{\bf j}}%
N\right) \right\rangle _{{\bf j}}  \nonumber \\
&=&(4\pi \hbar )^{-L}\left\langle e^{i2\pi x\wedge {\bf j}}\widehat{{\bf T}}%
_{\frac{{\bf j}}N}\right\rangle _{{\bf j}}.
\end{eqnarray}
To perform the average we use the periodicity of the torus translation (\ref
{eq:TTsim}), but we perform two loops around the torus, so that 
\begin{eqnarray}
\widehat{{\bf R}}_x^\chi &=&(4\pi \hbar )^{-L}\left( \frac 1{2N}\right)
^2\sum_{r=0}^{2N-1}\sum_{s=0}^{2N-1}\left\langle e^{-i2\pi N\left[ \left( 
\frac{{\bf j}}{2N}+\frac \chi N\right) \wedge 2{\bf k}+{\bf k}\widetilde{%
{\frak J}}{\bf k}\right] }e^{i2\pi x\wedge ({\bf j+}2{\bf k}N{\bf )}}%
\widehat{{\bf T}}_{\frac{{\bf j}}N}\right\rangle _{{\bf k}}  \nonumber \\
&=&\frac 1{2N}\sum_{r=0}^{2N-1}\sum_{s=0}^{2N-1}e^{-i2\pi x\wedge {\bf j}}%
\widehat{{\bf T}}_{\frac{{\bf j}}N}\left\langle e^{-i2\pi \left[ \left( \chi
-xN\right) \wedge 2{\bf k}\right] }\right\rangle _{{\bf k}}.
\end{eqnarray}
The ${\bf k}$ average is different from zero only if the point $x$ is such
that $x=x_{a,b}\equiv \frac 1N\left( 
\begin{array}{c}
a+\chi _p \\ 
b+\chi _q
\end{array}
\right) ,\,$ with $a\,$ and $b$ half integers.

In short,
\begin{eqnarray}
\widehat{{\bf 1}}_N^\chi \widehat{R}_x\widehat{{\bf 1}}_N^\chi &=&\left\{ 
\begin{tabular}{lll}
$\widehat{{\bf R}}_{x_{a,b}}^\chi $ & $\mbox{if }$ $x=\left( \frac{a+\chi _p}%
N,\frac{b+\chi _q}N\right) ,$ where $a$ and $b$ are half-integers &  \\ 
$0$ & otherwise & 
\end{tabular}
\right. \\
&=&\widehat{{\bf R}}_{x_{a,b}}^\chi \left\langle \delta \left( x-\frac{{\bf k%
}}{2N}+\frac \chi N\right) \right\rangle _{{\bf k}}=\widehat{R}_x\widehat{%
{\bf 1}}_N^\chi ,  \label{rtoro}
\end{eqnarray}
where 
\begin{equation}
\widehat{{\bf R}}_{x_{a,b}}^\chi =\frac
1{2N}\sum_{r=0}^{2N-1}\sum_{s=0}^{2N-1}e^{i2\pi Nx\wedge \xi }\widehat{{\bf T%
}}_\xi ^\chi  \label{eq:RRT}
\end{equation}
is the torus reflection on the center point $x_{a,b}$. The last equality in (%
\ref{rtoro}) holds because $\widehat{{\bf R}}_{x_{a,b}}^\chi $ commutes with 
$\widehat{{\bf 1}}_N^\chi .$ Again, we will usually omit the explicit $\chi 
$ dependence.

The construction of the reflection operators on the torus replaces the
Fourier transform (\ref{eq:rint}) by a Fourier sum on the torus translation
operators. The sums are to be taken over operators on one complete period;
the symmetry properties (\ref{eq:TTsim}) show that this period is obtained
with chords that perform two loops around the torus, i.e. , the period is
double that expected. So, although the basis of operators is formed with
chords that perform up to one loop around the torus in the Fourier sum, we
have to sum over chords that perform up to two loops. Thus, the basis
operators are summed twice, but with different Fourier phases.

In what follows, the subscripts $(a,b)$ for the discrete centers and $(r,s)$
for the lattice of chords are implicit and they will be explicitly written
only to avoid possible confusion. With the use of (\ref{eq:114}) we can
derive the following extensively used relations, 
\begin{equation}
\sum_{a=0}^{N-1/2}\sum_{b=0}^{N-1/2}e^{i2\pi Nx\wedge \xi }=(2N)^2\delta
_{r,0}^{(2N)}\delta _{s,0}^{(2N)},  \label{eq5}
\end{equation}
where all the sums over $a$ and $b$ are taken with step $\frac 12$, and 
\begin{equation}
\sum_{r=0}^{2N-1}\sum_{s=0}^{2N-1}e^{i2\pi Nx\wedge \xi }=(2N)^2\delta
_{b,0}^{(N)}\delta _{a,0}^{(N)}.  \label{eq6}
\end{equation}
Here $\delta _{b,0}^{(2N)}$ is a period-$2N$ Kronecker delta function.
Inserting (\ref{eq:TQ}) in (\ref{eq:RRT})we find the action of the
reflection operators on the Hilbert space:

\begin{equation}
\widehat{{\bf R}}_x|{\bf q}_n,N>=e^{i\frac{2\pi }N2(b-n)(a+\chi _p)}|{\bf q}%
_{2b-n},N>  \label{eq:RQ}
\end{equation}
and 
\begin{equation}
\widehat{{\bf R}}_x|{\bf p}_m,N>=e^{i\frac{2\pi }N2(a-m)(b+\chi _q)}|{\bf p}%
_{2a-m},N>.  \label{eq:RP}
\end{equation}
The unitarity of $\widehat{{\bf R}}_x$ is ensured by (\ref{eq:RQ}) and (\ref
{eq:RP}).

\begin{figure}[htb]
\centerline {\epsfxsize=3in  \epsffile{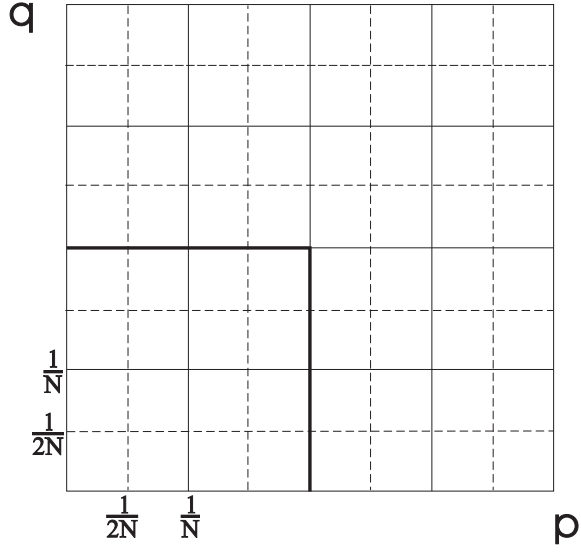} }
\caption{  The quantum Phase space QPS for $N=4$ (solid line). The Weyl
phase space WPS (doted line) is the grid of points $x$, centers of the
reflection operators in QPS. The area in the bold square is the
quarter-torus in which lie the centers $x,$ which label the basis of
operators ${\bf R}_x$.
}
\label{fig.3.2}
\end{figure}

We then see that $\widehat{{\bf R}}_x$ reflects the (QPS) web about the
point $x=(\frac{a+\chi _p}N,\frac{b+\chi _q}N)$. We need to include
half-integer values of $a$ and $b$ so that with a given $|{\bf q}_n,N>$ we
can span all ${\cal H}_N^\chi $ by applying different $\widehat{{\bf R}}_x$.
This is in complete agreement with the fact that the reflections that leave
invariant the web formed by the QPS must include half-integer values of $a$
and $b$, conferring on these half-integers a clear geometrical meaning. So
the centers of the reflections form a web whose spacing is half that of the
QPS, as shown in Fig.~\ref{fig.3.2}. Once more, the only operators that do not vanish on
projection to the torus are those that leave ${\cal H}_N^\chi $ invariant.
These correspond classically to those transformations that leave the QPS web
invariant.

The matrix elements of the reflection operators in the $|{\bf q}_n,N>$ basis
are 
\begin{equation}
<{\bf q}_m,N|\widehat{{\bf R}}_x|{\bf q}_n,N>=e^{i\frac{2\pi }N(m-n)(a+\chi
_q)}\delta _{m,2b-n}^{(N)}e^{i\frac{2\pi }Na(2b-n-m)}.  \label{eq:RQQ}
\end{equation}
From (\ref{eq:RQ}) we can see the symmetry properties of these operators, 
\begin{equation}
\widehat{{\bf R}}_{x+\frac{{\bf k}}2}=(-1)^{bk_p+ak_q+k_pk_qN}\widehat{{\bf R%
}}_x  \label{eq:symR}
\end{equation}
where ${\bf k}=(k_p,k_q)$ is a vector with integer components. It is
important to see that the domain of the variables $a$ and $b$ being integer
and half-integer values,  we have $(2N)^2$ different
reflection operators in the unit square. But the symmetry properties (\ref{eq:symR}) show that
only $N^2$ of them are independent, so we take the values of $a$ and $b$
that belong to $[0,\frac{N-1}2]$ ; this forms a complete set of independent
operators. That is, only one quarter of the torus is needed to define a
complete set of reflection operators. The values of $x=(\frac{a+\chi _p}N,%
\frac{b+\chi _q}N)$ generated by these values of $a$ and $b$ do not all
belong to the QPS; indeed we define here another space, the {\it Weyl phase
space}, WPS, formed by the support of $x$ this is shown by the bold face
area in Fig.~\ref{fig.3.2}. In the case where $N$ is odd, we will see later that WPS can
be defined such that it coincides with the QPS.

By the use of (\ref{eq5}) we can see that, 
\begin{equation}
\widehat{{\bf T}}_\xi =\frac
1{2N}\sum_{a=0}^{N-1/2}\sum_{b=0}^{N-1/2}e^{-i2\pi Nx\wedge \xi }\widehat{%
{\bf R}}_x,  \label{eq:tsumr}
\end{equation}
where we are again taking the sum with the indices running in an interval
twice as large as that needed to define a basis of operators. This is so
because (\ref{eq:symR}) implies that classically equivalent reflections,
through points diametrically opposed on any of the circuits of the torus,
are only equal up to a phase.

We now investigate the group or cocycle properties of the translations and
reflections defined in this section. It is important to note that the
transformations treated here are such that they leave the web formed by the
QPS invariant at the classical level, as well as the Hilbert space ${\cal H}%
_N^\chi $. With the help of (\ref{eq:tsumr}) and (\ref{eq:TT}), we obtain
the following properties for these operators, 
\begin{equation}
\widehat{{\bf R}}_x\widehat{{\bf T}}_\xi =\widehat{{\bf R}}_{x-\xi
/2}e^{-i2\pi Nx\wedge \xi },  \label{eq:RT}
\end{equation}
\begin{equation}
\widehat{{\bf T}}_\xi \widehat{{\bf R}}_x=\widehat{{\bf R}}_{x+\xi
/2}e^{-i2\pi Nx\wedge \xi },  \label{eq:TR}
\end{equation}
\begin{equation}
\widehat{{\bf R}}_{x_1}\widehat{{\bf R}}_{x_2}=\widehat{{\bf T}}%
_{2(x_2-x_1)}e^{i2\pi N(2x_1\wedge x_2)}.  \label{eq:RR}
\end{equation}
We then have the same cocycle properties as in the plane: (\ref{eq:tt})-(\ref
{eq:rr}). This is a consequence of the commutation of operator products with
projection (\ref{eq:abpt}) and will be of crucial importance when we derive
the properties of the center and chord representations on the torus. Note
that the characterization of the chords $\xi $ by integers and the centers $%
x $ by half-integers is respected by the group of operations above.

Another property which results from the last cocycle relation (\ref{eq:RR})
is 
\begin{equation}
\widehat{{\bf R}}_x\widehat{{\bf R}}_x=\widehat{{\bf 1}}_N^\chi ,
\end{equation}
in accordance with classical reflections. This means that 
\begin{equation}
\widehat{{\bf R}}_x^{-1}=\widehat{{\bf R}}_x^{\dag }=\widehat{{\bf R}}_x,
\end{equation}
that is, reflection operators on the torus are unitary and Hermitian.

It is important at this stage to examine the trace of these operators. Using
(\ref{eq:TQQ}) and (\ref{eq:114}), we have: 
\begin{equation}
{\bf Tr}(\widehat{{\bf T}}_\xi )=Ne^{i\frac{2\pi }N(\frac{rs}2+r\chi
_p-s\chi _p)}\delta _{r,0}^{(N)}\delta _{s,0}^{(N)}\equiv N\;e^{i\frac{2\pi }%
N(\frac{rs}2+r\chi _q-s\chi _p)}\;\delta _\xi ^{(N)}.  \label{eq:trT}
\end{equation}
For the trace of the reflection operators, we recall (\ref{eq:RQQ}), so

\begin{equation}
<{\bf q}_n,N|\widehat{{\bf R}}_x|{\bf q}_n,N>=\delta _{n,2b-n}^{(N)}e^{i%
\frac{2\pi }Na(2b-2n)}\ ,  \label{eq:Rnn}
\end{equation}
which is different from zero only if

\begin{equation}
n=b\qquad \mbox{mod}\left( \frac N2\right) .  \label{eq:nb}
\end{equation}
However, if $b$ is half-integer and $N$ is even, for example, there would be
no $n$ such that (\ref{eq:nb}) is satisfied. In general we can have up to 2
solutions of (\ref{eq:nb}) for $n\in \left[ 0,N-1\right] $, but they can
have different phase contributions in (\ref{eq:Rnn}). A careful inspection
leads to: 
\begin{eqnarray}
{\bf Tr}(\widehat{{\bf R}}_x) &=&f_N(x)=\frac
12(1+(-1)^{2a}+(-1)^{2b}+(-1)^{2a+2b+N})=  \nonumber  \label{eq:trRT1} \\
&=&\left\{ 
\begin{tabular}{ll}
$0$ & $\mbox{ if N is even and }a\mbox{ or }b\mbox{ semi-integers}$ \\ 
$2$ & $\mbox{ if N is even and }a\mbox{ and }b\mbox{ integers}$ \\ 
$1$ & $\mbox{if N is odd and }a\mbox{ or }b\mbox{ integers}$ \\ 
$-1$ & $\mbox{ if N is odd and }a\mbox{ and }b\mbox{ semi-integers}$%
\end{tabular}
\right.  \label{eq:trRT}
\end{eqnarray}
The importance of this result for the following theory calls for some
intuitive explanation in terms of the reflections of the discrete periodic
lattice. As in the plane case, we can relate ${\bf Tr}(\widehat{{\bf R}}_x)$
to the number of fixed points of the corresponding classical map. Indeed,
for $N$ odd there is always a single fixed point, agreeing with the modulus
of (\ref{eq:trRT}). If $N$ is even, there will only be fixed points if $x$
is characterized by integer numbers ($a,b$), in which case there are two.

\subsection{Operators and their Symbols}

Once we have defined the reflection and translation operators, we can
decompose any operator as their linear combination. To construct the chord,
or translation representation of an operator, we express any operator as a
linear combination of translations. To have a complete basis, we need just $%
N^2$ operators, so that $r$ and $s$ run from $0$ to $N-1$. The chords $\xi
=(\frac rN,\frac sN)$ having this property are said to belong to the
fundamental domain. The other translation operators are obtained from these
through the symmetry properties; that is, the fundamental translations are
those which have chords smaller than one loop around any of the irreducible
circuits of the torus in a given direction. The chord representation of an
operator is defined as

\begin{equation}
{\bf A}(\xi )\equiv {\bf Tr}\left( \widehat{{\bf A}}\widehat{{\bf T}}_{-\xi
}\right) .  \label{eq:Acor}
\end{equation}
From the symbol, we recover the operator: 
\begin{equation}
\widehat{{\bf A}}=\frac 1N\sum_{r,s=0}^{N-1}{\bf A}(\xi )\widehat{{\bf T}}%
_\xi \equiv \frac 1N\sum_\xi {\bf A}(\xi )\widehat{{\bf T}}_\xi .
\label{eq:corA}
\end{equation}

Although, to recover the operator we only need the symbol defined in the
fundamental domain (i.e. $r,s$ in $[0,N-1]$), (\ref{eq:Acor}) can be used to
extend the definition of the symbol for $r$ and $s$ running among all
integer numbers. Of course, these will not be independent of the symbols in
the fundamental domain and, from the symmetry properties of $\widehat{{\bf T}%
}(\xi )$ (\ref{eq:TTsim}), we see that ${\bf A}(\xi )$ satisfies 
\begin{equation}
{\bf A}(\xi +{\bf k})=(-1)^{sk_p+rk_q+k_pk_qN}e^{i2\pi (k_p\chi _q-k_q\chi
_p)}{\bf A}(\xi )=e^{-i2\pi N\left[ \left( \frac \xi 2+\frac \chi N\right)
\wedge {\bf k}+\frac 14{\bf k}\widetilde{{\frak J}}{\bf k}\right] }{\bf A}%
(\xi ),  \label{eq:Acorsim}
\end{equation}
where ${\bf k}=(k_p,k_q)$ is a vector with integer components denoting
chords that perform respectively $k_p$ and $k_q$ loops around the
irreducible circuits of the torus. This is an important consequence of the
fact that the symmetry properties of the symbol of operators are the same as
those of the basis operators used to generate this symbol.

We now expand the operators in term of reflections; this is the center or
Weyl representation. It is important to recall that we must take values of $%
a $ and $b$ that belong to $[0,\frac{N-1}2]$, that is, only one quarter of
the torus is needed to define a complete set of reflection operators . The
values of $x=(\frac{a+\chi _p}N,\frac{b+\chi _q}N)$ generated by these
values of $a$ and $b$ define the Weyl phase space ,WPS, shown by the bold
face area in Fig.~\ref{fig.3.2}.

We define the center symbol of an operator $\widehat{{\bf A}}${\bf \ }such
that,

\begin{equation}
{\bf A}(x)\equiv {\bf Tr}\left( \widehat{{\bf A}}\widehat{{\bf R}}_x\right) .
\label{eq:Acen}
\end{equation}
From the symbol, we recover the operator through 
\begin{equation}
\widehat{{\bf A}}=\frac 1N\sum_{a,b=0}^{\frac{N-1}2}\widehat{{\bf R}}_x{\bf A%
}(x)\equiv \frac 1N\sum_x\widehat{{\bf R}}_x{\bf A}(x).  \label{eq:cenA}
\end{equation}
The symmetry properties of $\widehat{{\bf R}}_x$ (\ref{eq:symR}) imply 
\begin{equation}
{\bf A}(x+\frac{{\bf k}}2)=(-1)^{bk_p+ak_q+k_pk_qN}{\bf A}(x),
\label{eq:Acensim}
\end{equation}
for any vector ${\bf k}=(k_p,k_q)$ with integer components. This result had
already been obtained by Hannay and Berry \cite{hanay} for the Wigner
function and we see here that it is general for any Weyl symbol on the torus.

As in the plane case, we derive some important properties of the
translations and Weyl symbols. Notice first that:

\begin{equation}
{\bf R}_x(\xi )={\bf Tr}(\widehat{{\bf R}}_x\widehat{{\bf T}}_{-\xi })={\bf %
Tr}(\widehat{{\bf R}}_{x+\xi /2})e^{i2\pi Nx\wedge \xi }=f_N(x+\xi
/2)e^{i2\pi Nx\wedge \xi }
\end{equation}
and

\begin{equation}
{\bf T}_\xi (x)={\bf Tr}(\widehat{{\bf T}}_\xi \widehat{{\bf R}}_x)={\bf Tr}(%
\widehat{{\bf R}}_{x+\xi /2})e^{-i2\pi Nx\wedge \xi }=f_N(x+\xi /2)e^{-i2\pi
Nx\wedge \xi }.
\end{equation}

The trace is now obtained as

\begin{eqnarray}
{\bf Tr}\left( \widehat{{\bf A}}\right) &=&{\bf A}(\xi =0)  \label{eq:trAcor}
\\
&=&\sum_x{\bf A}(x)f_N(x)=\frac 12\sum_{a,b=0}^{N-\frac 12}{\bf A}(x).
\end{eqnarray}
In the last equality we use the fact that the Weyl symbols for the entire
torus are obtained from those of a quarter of it through the symmetry
relations (\ref{eq:Acensim}) and the definition of $f_N(x)$ (\ref{eq:trRT1}).

The representation of the identity on the torus Hilbert space ${\cal H}%
_N^\chi $ has now the form:

\begin{equation}
{\bf 1}_N^\chi (x)=f_N(x)\ \ \ \mbox{ and }\ \ \ {\bf 1}_N^\chi (\xi
)=N\;\delta _\xi ^{(N)}.
\end{equation}

Hermitian operators are associated to the observables of the system and, in
particular, the Hamiltonian generates the dynamics. Defined as $\widehat{%
{\bf H}}=\widehat{{\bf H}}^{\dag }$, we obtain
\begin{equation}
{\bf H}^{\dag }(\xi )=[{\bf H}(-\xi )]^{*}\ \ \ \mbox{and}\ \ \ {\bf H}%
^{\dag }(x)=[{\bf H}(x)]^{*}\ ,
\end{equation}
just as for the plane case \cite{ozrep}.

The role played in the plane case by the Fourier transform will be taken by
the finite Fourier transform, since it allows us to exchange chords and
centers as well as to change from center or chord to the position
representation. But there are some small differences due to the $f_N(x)$
factors peculiar to the torus. Thus, in the exchange of centers and chords
we have, 
\begin{eqnarray}
{\bf A}(\xi ) &=&{\bf Tr}\left( \frac 1N\sum_x\widehat{{\bf R}}_x{\bf A}(x)%
\widehat{{\bf T}}_{-\xi }\right) =\frac 1N\sum_x{\bf A}(x){\bf Tr}(\widehat{%
{\bf R}}_{x+\xi /2})e^{i2\pi Nx\wedge \xi }  \nonumber \\
&=&\frac 1N\sum_x{\bf A}(x)f_N(x+\xi /2)e^{i2\pi Nx\wedge \xi },
\end{eqnarray}
whereas 
\begin{equation}
{\bf A}(x)=\frac 1N\sum_\xi {\bf A}(\xi )f_N(x+\xi /2)e^{-i2\pi Nx\wedge \xi
}.
\end{equation}

Using (\ref{eq:TQQ}) and (\ref{eq:Acorsim}) we obtain the position
representation of an operator $\widehat{{\bf A}}$%
\begin{eqnarray}
<{\bf q}_m,N|\widehat{{\bf A}}|{\bf q}_n,N> &=&\frac 1N\sum_\xi {\bf A}(\xi
)<{\bf q}_m,N|\widehat{{\bf T}}_\xi |{\bf q}_n,N>  \nonumber \\
&=&\frac 1N\sum_{r,s=0}^{N-1}{\bf A}(\xi _{r,s})e^{-i\frac{2\pi }Nr(\frac{m+n%
}2+\chi _q)}\delta _{m,n+s}^{(N)}e^{i\frac{2\pi }N(\frac r2+\chi _p)(m-n-s)}
\nonumber \\
&=&\frac 1N\sum_{r=0}^{N-1}{\bf A}(\xi _{r,m-n})e^{-i\frac{2\pi }Nr(\frac{m+n%
}2+\chi _q)}.  \label{eq:AQsig}
\end{eqnarray}
Note that in this last equation we are employing chords that may not belong
to the fundamental domain; that is $m-n$ may not belong to $[0,N-1]$.
However, the symbol for this chord is well defined through (\ref{eq:Acor}).
If we restrict ourselves to chords that belong to the fundamental domain, we
then have a supplementary $e^{i\frac{2\pi }N(\frac r2+\chi _p)(m-n)}$ phase
factor in the last sum. This kind of difficulty may appear in the following
formulae, but, by allowing the indices to run over all integer numbers, the
formulae become indeed much simpler, as is the case for (\ref{eq:AQsig}).

Using the position representation 
\begin{equation}
\widehat{{\bf A}}=\sum_{m,n=0}^{N-1}|{\bf q}_m,N><{\bf q}_m,N|\widehat{{\bf A%
}}|{\bf q}_n,N><{\bf q}_n,N|,
\end{equation}
we retrieve the chord representation as 
\begin{eqnarray}
{\bf A}(\xi ) &=&{\bf Tr}\left( \sum_{m,n=0}^{N-1}<{\bf q}_m,N|\widehat{{\bf %
A}}|{\bf q}_n,N>|{\bf q}_m,N><{\bf q}_n,N|\widehat{{\bf T}}_{-\xi }\right) 
\nonumber \\
&=&\sum_{m,n=0}^{N-1}<{\bf q}_m,N|\widehat{{\bf A}}|{\bf q}_n,N>e^{-i\frac{%
2\pi }Nr(\frac{m+n}2+\chi _q)}\delta _{m,n+s}^{(N)}e^{i\frac{2\pi }N(\frac
r2+\chi _q)(m-n-s)}  \nonumber \\
&=&\sum_{n=0}^{N-1}<{\bf q}_{n+s},N|\widehat{{\bf A}}|{\bf q}_n,N>e^{-i\frac{%
2\pi }Nr(n+\frac s2+\chi _q)}.  \label{AcorQ}
\end{eqnarray}
Using (\ref{eq:RQQ}) and (\ref{eq:Acensim}) we exchange the coordinate and
the center representation: 
\begin{eqnarray}
<{\bf q}_m,N|\widehat{{\bf A}}|{\bf q}_n,N> &=&\frac 1N\sum_x{\bf A}(x)<{\bf %
q}_m,N|\widehat{{\bf R}}_x|{\bf q}_n,N>  \nonumber \\
&=&\frac 1N\sum_{a,b=0}^{\frac{N-1}2}{\bf A}(x_{a,b})e^{i\frac{2\pi }%
N2(b-n)(a+\chi _p)}\delta _{m,2b-n}^{(N)}e^{i\frac{2\pi }N2\chi _p(m-2b+n)} 
\nonumber \\
&=&\frac 1N\sum_{a=0}^{\frac{N-1}2}{\bf A}(x_{a,\frac{m+n}2})e^{i\frac{2\pi }%
N(m-n)(a+\chi _p)}.  \label{eq:AQx}
\end{eqnarray}
Note that in this last formula we are taking the center point $x$ that does
not belong to the fundamental domain (i.e. $\frac{m+n}2$ may not belong to $%
[0,\frac{N-1}2]$ ). We recover the center representation through 
\begin{eqnarray}
{\bf A}(x) &=&{\bf Tr}\left( \sum_{m,n=0}^{N-1}<{\bf q}_m,N|\widehat{{\bf A}}%
|{\bf q}_n,N>|{\bf q}_m,N><{\bf q}_n,N|\widehat{{\bf R}}_x\right)  \nonumber
\\
&=&\sum_{m,n=0}^{N-1}<{\bf q}_m,N|\widehat{{\bf A}}|{\bf q}_n,N>e^{i\frac{%
2\pi }N(m-n)(a+\chi _p)}\delta _{m,2b-n}^{(N)}e^{i\frac{2\pi }Na(2b-n-m)} 
\nonumber \\
&=&\sum_{n=0}^{N-1}<{\bf q}_{2b-n},N|\widehat{{\bf A}}|{\bf q}_n,N>e^{i\frac{%
2\pi }N2(b-n)(a+\chi _p)}.  \label{AxQ}
\end{eqnarray}

It would be possible to define the chord and the Weyl representations by
equations (\ref{AcorQ}) and (\ref{AxQ}) respectively. However, the
geometrical structure, the role of the translations and reflection operators
and the relation to the plane theory would then be relegated to curiosities.

\subsection{Symbols of the product of operators}

We now derive the product law of the symbols of the operators in these
representations. Let us start with the chord representation (\ref{eq:corA}).
For this purpose we use ,(\ref{eq:TT}) and (\ref{eq:trT}) to obtain 
\begin{eqnarray}
{\bf AB}(\xi ) &=&{\bf Tr}\left( \widehat{{\bf A}}\widehat{{\bf B}}\widehat{%
{\bf T}}_{-\xi }\right) =(\frac 1N)^2\sum_{\xi _1,\xi _2}{\bf A}(\xi _1){\bf %
B}(\xi _2){\bf Tr}\left( \widehat{{\bf T}}_{\xi _1}\widehat{{\bf T}}_{\xi _2}%
\widehat{{\bf T}}_{-\xi }\right)  \nonumber \\
&=&(\frac 1N)^2\sum_{\xi _1,\xi _2}{\bf A}(\xi _1){\bf B}(\xi _2)Ne^{i2\pi
ND_4(\xi _1,\xi _2,-\xi )}{\bf Tr}\left( \widehat{{\bf T}}_{\xi _1+\xi
_2-\xi }\right)  \nonumber \\
&=&(\frac 1N)\sum_{\xi _1}{\bf A}(\xi _1){\bf B}(\xi -\xi _1)e^{i2\pi
ND_3(\xi _1,-\xi )},  \label{eq:ABcor}
\end{eqnarray}
where we allow chords $\xi _2=\xi -\xi _1$ not to be in the fundamental
domain. Let us now take the trace of the product; inserting (\ref{eq:trAcor}%
) in (\ref{eq:ABcor}) leads to 
\begin{equation}
{\bf Tr}\left( {\bf AB}\right) =(\frac 1N)\sum_{\xi _1}{\bf A}(\xi _1){\bf B}%
(-\xi _1).
\end{equation}
The generalization of (\ref{eq:ABcor}) for the product of an arbitrary
number of operators is 
\begin{eqnarray}
{\bf A}_n...{\bf A}_1(\xi ) &=&(\frac 1N)^{n-1}\sum_{\xi _1...\xi _{n-1}}%
{\bf A}_1(\xi _1)...  \nonumber \\
&&\!{\bf A}_{n-1}(\xi _{n-1}){\bf A}_n(\xi -\sum_{i=1}^{n-1}\xi _i)\exp
\left[ -i2\pi ND_{n+1}(\xi _1,...,\xi _{n-1},-\xi )\right] .  \label{prodtr}
\end{eqnarray}
Thus, the product rule for the chords is obtained from that in the plane by
simply substituting the integral in (\ref{eq:ancor}) by the corresponding
sum.

\begin{figure}[htb]
\centerline {\epsfxsize=5in  \epsffile{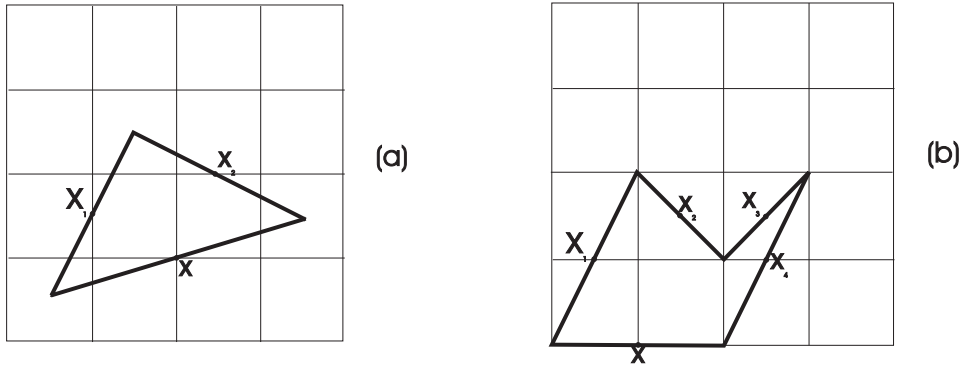} }
\caption{   Two examples of polygons displaying the uniform nature of
the vertices of the polygon. (a) all vertices lie on half integers: $f_N=0$
or $-1$ for $N$ respectively even or odd (b) the vertices lie on integers: $%
f_N=2 $ or $1$ for $N$ respectively even or odd.
}
\label{fig.3.3}
\end{figure}

For the center symbol (\ref{eq:Acen}) the trace of the product is obtained
using (\ref{eq:RR}) and (\ref{eq:trT}): 
\begin{eqnarray}
{\bf Tr}\left( {\bf AB}\right) &=&(\frac 1N)^2\sum_{x_1,x_2}{\bf A}(x_2){\bf %
B}(x_1){\bf Tr}\left( \widehat{{\bf R}}_{x_2}\widehat{{\bf R}}_{x_1}\right) 
\nonumber \\
&=&(\frac 1N)^2\sum_{x_1,x_2}{\bf A}(x_2){\bf B}(x_1)e^{i2\pi N(2x_1\wedge
x_2)}{\bf Tr}\left( \widehat{{\bf T}}_{2(x_2-x_1)}\right)  \nonumber \\
&=&(\frac 1N)\sum_{x_1}{\bf A}(x_1){\bf B}(x_1).
\end{eqnarray}
We will now derive the full product properties in the center representation (%
\ref{eq:cenA}); with the help of the group properties (\ref{eq:RR}), (\ref
{eq:RT}) and (\ref{eq:trRT}) we have 
\begin{eqnarray}
{\bf AB}(x) &=&{\bf Tr}\left( \widehat{{\bf A}}\widehat{{\bf B}}\widehat{%
{\bf R}}_x\right) =(\frac 1N)^2\sum_{x_1,x_2}{\bf A}(x_2){\bf B}(x_1){\bf Tr}%
\left( \widehat{{\bf R}}_{x_2}\widehat{{\bf R}}_{x_1}\widehat{{\bf R}}%
_x\right)  \nonumber \\
&=&(\frac 1N)^2\sum_{x_1,x_2}{\bf A}(x_2){\bf B}(x_1)e^{i2\pi N2(x_1\wedge
x_2+x_2\wedge x+x\wedge x_1)}{\bf Tr}\left( \widehat{{\bf R}}%
_{x+x_2-x_1}\right)  \nonumber \\
&=&(\frac 1N)^2\sum_{x_1,x_2}{\bf A}(x_2){\bf B}(x_1)e^{i2\pi N\Delta
_3(x,x_1,x_2)}f_N(x+x_2-x_1),  \label{eq:ABcen}
\end{eqnarray}
where the symplectic area of the triangle $\Delta _3(x,x_1,x_2)$ was defined
in section 3. Note that the sides of these triangles must be integer vectors
in this case, because the symmetry of each side about its center implies
that all the corners will be of the same type regardless of whether either $%
a $ or $b$ are integer or half-integer. The argument of the function $f_N$
defined in (\ref{eq:trRT}) can thus be any corner of the triangle as shown
in Fig.~\ref{fig.3.3}(a). We thus find that the reflection properties of the QPS lead to
a more complex product rule than for the plane (\ref{eq:aacen}).

The generalization for the product of $2n$ operators is 
\begin{eqnarray}
&&{\bf A}_{2n}...{\bf A}_1(x)=(\frac 1N)^{2n}\sum_{x_1...x_{2n}}{\bf A}%
_{2n}(x_{2n})  \nonumber \\
&&{\bf ...A}_1(x_1)e^{i2\pi N\Delta
_{2n+1}(x,x_{2n},...,x_1)}f_N(x+\sum_{j=1}^{2n}(-1)^jx_j),  \label{3.77}
\end{eqnarray}
where again the argument of $f_N$ is any corner of the polygon whose centers
are $x,x_{2n},...,x_1$ (see an example in Fig.~\ref{fig.3.3}(b)). For an odd number of
operators we just choose $\widehat{{\bf A}}_1=\widehat{{\bf 1}}$, that is $%
{\bf A}_1(x_1)=f_N(x_1)$ in (\ref{3.77}).

The product laws are the main result of this section. In contrast with the
Weyl-like representation obtained by Galleti and Toledo Pisa \cite{galeti2},
we only need half the number of sums (including the implicit sums in the trace of their formula (21)). Kaperskovitz and Peev \cite{kaperpeev}
also have a Weyl-like representation, but only for the case of $N$ even.
They perform products of 2 operators and the product law that they obtain is
very similar to ours, although their result is not compatible with our geometrical interpretation, because we need half-integer vectors to completely describe the reflections of QPS. Most important is the fact that our formalism prescribes the product of an arbitrary even number of operators, just as for the plane, whereas previous results could only cope explicitly with the product of two operators at a time.

\subsection{Weyl representation in QPS}

If $N$ is odd, we can redefine the WPS so that it coincides with the QPS.
For this purpose we define $X=(\frac{\alpha +\chi _p}N,\frac{\beta +\chi _q}%
N)$ so that ,

\begin{eqnarray}
\alpha &=&a+j\frac N2\qquad \mbox{ where }j=\left\{ 
\mbox{\begin{tabular}{ll}
$0$ & $\quad \mbox{ if }a\mbox{ is integer}$ \\ 
$1$ & $\quad \mbox{ otherwise}$\end{tabular}
}\right. \\
\beta &=&b+k\frac N2\qquad \mbox{ where }k=\left\{ 
\mbox{\begin{tabular}{ll}
$0$ & $\quad \mbox{ if }b\mbox{ is integer}$ \\ 
$1$ & $\quad \mbox{ otherwise}$\end{tabular}
}\right.
\end{eqnarray}
We then have that $\alpha $ and $\beta $ are integers for the case were $N$
is odd. In other words for any $x$ there is a point $X$ such that 
\begin{equation}
X=\frac 1N\left( 
\begin{array}{c}
\alpha \\ 
\beta
\end{array}
\right) +\frac \chi N=x+\frac 12{\bf n}  \label{xentero}
\end{equation}
with ${\bf n}$ an integer vectors. If $N$ is even, the $\alpha $ and $\beta $
will have the same character ( integer or half-integer) as $a$ and $b$, so
we cannot recover the QPS. In the rest of this section we will then restrict
ourselves to the case where $N$ is odd. The symmetry relation (\ref{eq:symR}%
) shows then that 
\begin{equation}
\widehat{{\bf R}}_X=(-1)^{(2jb+2ja+jk)}\widehat{{\bf R}}_x,
\end{equation}
and with the use of (\ref{eq:trRT}) we have

\begin{equation}
{\bf Tr}(\widehat{{\bf R}}_X)=1.
\end{equation}

We now see that letting $a$ and $b$ run over the half-integers in $[0,\frac{%
N-1}2]$, we then have $\alpha $ and $\beta $ integers in $[0,N-1]$ and we
recover the QPS. For this space we will now have a new Weyl representation 
\begin{equation}
{\bf A}(X)={\bf Tr}\left( \widehat{{\bf A}}\widehat{{\bf R}}_X\right) ,
\end{equation}
from which we recover the operator: 
\begin{equation}
\widehat{{\bf A}}=\frac 1N\sum_{\alpha ,\beta =0}^{N-1}\widehat{{\bf R}}_X%
{\bf A}(X_{\alpha ,\beta })\equiv \frac 1N\sum_X\widehat{{\bf R}}_X{\bf A}%
(X).
\end{equation}

We will now examine the properties of the product in this representation: 
\begin{eqnarray}
{\bf AB}(X) &=&{\bf Tr}\left( \widehat{{\bf A}}\widehat{{\bf B}}\widehat{%
{\bf R}}_{X_{\alpha ,\beta }}\right) =(\frac 1N)^2\sum_{X_1,X_2}{\bf A}(X_2)%
{\bf B}(X_1){\bf Tr}\left( \widehat{{\bf R}}_{X_2}\widehat{{\bf R}}_{X_1}%
\widehat{{\bf R}}_X\right)  \nonumber \\
&&(\frac 1N)^2\sum_{X_1,X_2}{\bf A}(X_2){\bf B}(X_1)e^{i2\pi N2(X_1\wedge
X_2-X_2\wedge X+X\wedge X_1)}{\bf Tr}\left( \widehat{{\bf R}}%
_{X+X_2-X_1}\right)  \nonumber \\
&&(\frac 1N)^2\sum_{X_1,X_2}{\bf A}(X_2){\bf B}(X_1)e^{i2\pi N\Delta
_3(X,X_1,X_2)}.  \label{eq:ABinter}
\end{eqnarray}
This last expression is very similar to the general case described in the
previous section, but slightly simplified by the absence of the $f_N$ term,
in close analogy to the plane formalism. For the product of $2n$ operators
this generalizes to

\begin{equation}
{\bf A}_{2n}...{\bf A}_1(X)=(\frac 1N)^{2n}\sum_{X_1...X_{2n}}{\bf A}%
_{2n}(X_{2n}){\bf ...A}_1(X_1)e^{i2\pi N\Delta _{2n+1}(X,X_{2n,...},X_1)}.
\end{equation}
The absence of the $f_N$ factor in these simplified formulae may be
understood from the fact that a polygon whose centers all lie on an integer
lattice always has corners on the same lattice. Hence, $f_N$ is always unity
for all corners if $N$ is odd.

In the same way, for $N$ odd, we can perform a transformation, similar to (%
\ref{xentero}), to a set of chords $\tilde{\xi}$ that are even multiples of $%
\frac 1N$. This set of chords will be complete if we now allow them to
perform up to two loops around the circuits of the torus. This scheme can be
generalized to perform quantization on centers or chords that are multiples
of $\frac \phi N$ only if $2\phi $ and $N$ are coprime numbers. Then, the
chord (or center) will be supported by a lattice of spacing $\frac \phi N$
and length $\phi $. This transformation will have importance for cat maps
and will be studied in more details in reference \cite{loxocat}.

\subsection{Relation between symbols}

There are many ways to represent a given operator $\widehat{A}$ that acts
on the Banach space of the plane ${\cal H}_{{\Bbb R}}$. Among the different
representations, the center and chord symbols are of special interest in
this work. Projecting the operator $\widehat{A}$ onto the torus Hilbert
space ${\cal H}_N^\chi $ through $\widehat{{\bf A}}=\widehat{{\bf 1}}_N^\chi 
\widehat{A}\widehat{{\bf 1}}_N^\chi $, it can be represented in terms of
torus translations or reflections. We shall now show how the symbols on the
torus can be obtained from their counterparts on the plane.

Starting with the chord representation, we calculate the torus symbol at
points $\xi =\left( \frac rN,\frac sN\right) $. From the fact that $\widehat{%
{\bf T}}_{-\xi }$ and $\widehat{{\bf 1}}_N^\chi $ commute, we have 
\begin{equation}
{\bf A}(\xi )={\bf Tr}\left( \widehat{{\bf 1}}_N^\chi \widehat{A}\widehat{%
{\bf 1}}_N^\chi \widehat{{\bf T}}_{-\xi }\right) ={\bf Tr}\left( \widehat{%
{\bf 1}}_N^\chi \widehat{A}\widehat{T}_{-\xi }\widehat{{\bf 1}}_N^\chi
\right) ={\bf Tr}\left( \widehat{A}\widehat{T}_{-\xi }\widehat{{\bf 1}}%
_N^\chi \right) .
\end{equation}
Then, we express the operator $\widehat{A}$ in terms of translations (\ref
{eq:acor}) and use the group properties of the translation operators (\ref
{eq:tt}) to obtain 
\begin{equation}
{\bf A}(\xi )=\int \frac{d\xi _1}{(2\pi \hbar )}\ A(\xi _1)e^{\frac i{2\hbar
}\xi _1\wedge \xi }{\bf Tr}\left( \widehat{T}_{\xi _1-\xi }\widehat{{\bf 1}}%
_N^\chi \right) .
\end{equation}
We now use the projection properties of the translation operators on the
torus (\ref{projt}), so that 
\begin{equation}
{\bf A}(\xi )=\int \frac{d\xi _1}{(2\pi \hbar )}\ A(\xi _1)e^{\frac i{2\hbar
}\xi _1\wedge \xi }{\bf Tr}\left( \widehat{{\bf T}}_{\xi _1-\xi }\right)
\left\langle \delta \left( \xi _1-\xi -\frac{{\bf k}}N\right) \right\rangle
_{{\bf k}}.
\end{equation}
Performing the integral, with the help of the trace properties (\ref{eq:trT}%
), we obtain 
\begin{eqnarray}
{\bf A}(\xi ) &=&\left\langle (-1)^{sk_p+rk_q+k_pk_qN}e^{i2\pi (k_p\chi
_q-k_q\chi _p)}A\left( \frac rN+k_p,\frac sN+k_q\right) \right\rangle _{{\bf %
k}}=  \nonumber \\
&&\left\langle e^{i2\pi N\left[ \left( \frac \xi 2-\frac \chi N\right)
\wedge {\bf k+}\frac 14{\bf k}\widetilde{{\frak J}}{\bf k}\right] }A\left(
\xi +{\bf k}\right) \right\rangle _{{\bf k}},  \label{eq:Acorprom}
\end{eqnarray}
where the $(2L)$-dimensional vectors ${\bf k}$ have integer components. Note
that we have to perform a phase weighted average on equivalent points to
obtain the symbol on the torus. This is similar to the way Hannay and Berry
quantize the cat map\cite{hanay} in the coordinate representation.

We now proceed in a similar manner to derive the symbols in the center
representation at the points $x=(\frac{a+\chi _p}N,\frac{b+\chi _q}N)$.
Using the commutation of $\widehat{{\bf R}}_x$ with $\widehat{{\bf 1}}%
_N^\chi $ and (\ref{eq:acor}), we have 
\begin{equation}
{\bf A}(x)={\bf Tr}\left( \widehat{{\bf 1}}_N^\chi \widehat{A}\widehat{{\bf R%
}}_x\right) ={\bf Tr}\left( \widehat{{\bf 1}}_N^\chi \int \frac{dy}{(\pi
\hbar )}\ A(y)\hat{R}_y\widehat{R}_x\widehat{{\bf 1}}_N^\chi \right) ,
\end{equation}
which combined with the cocycle properties (\ref{eq:rr}), becomes 
\begin{equation}
{\bf A}(x)={\bf Tr}\left( \widehat{{\bf 1}}_N^\chi \int \frac{dy}{(\pi \hbar
)}\ A(y)e^{\frac i\hbar 2y\wedge x}\widehat{T}_{2(x-y)}\widehat{{\bf 1}}%
_N^\chi \right) .
\end{equation}
Projecting the translations on the torus (\ref{projt}), we have 
\begin{equation}
{\bf A}(x)=\int \frac{dy}{(\pi \hbar )}\ A(y)e^{\frac i\hbar 2y\wedge x}{\bf %
Tr}\left( \widehat{{\bf T}}_{2(x-y)}\right) \left\langle \delta \left(
2(x-y)-\frac{{\bf k}}N\right) \right\rangle _{{\bf k}},
\end{equation}
so that performing the integral we obtain, 
\begin{eqnarray}
{\bf A}(x) &=&\left\langle e^{i2\pi (ak_q-bk_p+\frac N2k_qk_p)}A\left( \frac{%
a+\chi _p}N+\frac{k_p}2,\frac{b+\chi _q}N+\frac{k_q}2\right) \right\rangle _{%
{\bf k}} =  \nonumber \\
& &\left\langle e^{i2\pi N(\left[ \left( x-\frac \chi N\right) \wedge {\bf k+%
}\frac 14{\bf k}\widetilde{{\frak J}}{\bf k}\right] )}A\left( x+\frac{{\bf k}%
}2\right) \right\rangle _{{\bf k}},  \label{eq:Axplator}
\end{eqnarray}
with the help of the trace properties (\ref{eq:trT}).

We again have a phase weighted average on equivalent points, this is a
general feature of any representation of projected operators ; only the
phase will depend on the specific representation we are taking. An important
feature of the center representation is that the phases do not have any
dependence on the $\chi $ parameters of the quantization; this is best seen
in (\ref{eq:Axplator}). Note also that comparing (\ref{eq:Acorprom}) and (\ref
{eq:Axplator}) with (\ref{eq:Acorsim}) and (\ref{eq:Acensim}) respectively,
 the phases are a consequence of the periodicity conditions of the
symbols. It is important to note that if $\widehat{A}$ and $\widehat{{\bf 1}}%
_N^\chi $ commute the restriction of $\widehat{A}$ on the Hilbert space $%
{\cal H}_N^\chi $ denotes an automorphism. Indeed, the commutation of $%
\widehat{A}$ and $\widehat{{\bf 1}}_N^\chi $ implies that the symbols $A(x)$
and $A(\xi )$ are periodic functions. Otherwise the average defined in (\ref
{eq:Acorprom}) and (\ref{eq:Axplator}) may not exist, it may happen that the
projected operator $\widehat{{\bf A}}=\widehat{{\bf 1}}_N^\chi \widehat{A}%
\widehat{{\bf 1}}_N^\chi =0.$

\subsection{Symplectic invariance}

At the end of section 3, we remarked that the center and the chord
representations in the plane are invariant with respect to the quantum
equivalents of linear canonical transformations, or symplectic
transformations $x^{\prime }=Mx.$ The transformations for which the
symplectic matrix $M$ is made up of integers are known colloquially as {\it %
cat maps}. These have the property that they leave invariant the unit torus.
Because of the commutation of operator products with projection from the
plane to the torus, the effect of a similarity transformation $\widehat{{\bf %
A}}\rightarrow \widehat{{\bf U}}_M\widehat{{\bf A}}\widehat{{\bf U}}_M^{-1}$
performed by a quantized cat map on any operator defined on the torus will
be purely classical in the center or the chord representations:

\begin{equation}
{\bf A}(x)\rightarrow {\bf A}(Mx)\quad \mbox{and\quad }{\bf A}(\xi
)\rightarrow {\bf A}(M\xi ) .  \label{cat1}
\end{equation}

Evidently, the matrix $M=1$ is a cat map; the product of cat maps is also a
cat map, as is the inverse of a cat map. It follows that the set of all cat
maps forms a subgroup of the symplectic transformations, which we will refer
to as the {\it feline group}. Likewise, relations (\ref{cat1}) indicate the
feline invariance of the chord and center representations.

In a companion paper \cite{loxocat} we use the chord and center
representations to study the properties of quantum cat maps of more than one
degree of freedom. This extends previous work by Hannay and Berry \cite
{hanay} and Keating \cite{keat1}, \cite{keat2} on two-dimensional cat maps.
For completeness, we will just note that the symbol corresponding to $%
\widehat{{\bf U}}_M$ is 
\begin{equation}
{\bf U}_M(X)=\frac 1{\sqrt{N^L}}\exp \left( \frac i\hbar XBX\right) ,
\label{catx}
\end{equation}
whereas the chord symbol is 
\begin{equation}
{\bf U}_M(\tilde{\xi})=\frac 1{\sqrt{N^L}}\exp \left( -\frac i\hbar \tilde{%
\xi}\beta \tilde{\xi}\right) .  \label{catcor}
\end{equation}
For quantization performed on $\chi =0$, with $N$ an odd integer, $B$ and $%
\beta $ are integer symmetric matrices and the chords $\tilde{\xi}=\frac
2N(r,s)$ (we perform quantization on a set of chords that are multiple of $%
\frac 2N$). The symmetric matrices $B$ and $\beta $ in the above quadratic
forms define the Cayley parametrization of the symplectic matrix $M:$%
\begin{equation}
M=\frac{1-{\frak J}B}{1+{\frak J}B}=\frac{1+{\frak J}\beta }{1-{\frak J}%
\beta }.
\end{equation}

In general, The Cayley matrices for a cat map may not be integer, leading to
less transparent relation between the symplectic classical matrix $M$ and
the symmetric matrices $B$ and $\beta $ in (\ref{catx}) and (\ref{catcor}).
They depend on Gaussian sums, as presented in reference \cite{loxocat}.

\section{Hamiltonians on the Torus and Path Integrals}

\setcounter{equation}{0}

We will now treat Hamiltonian systems on the torus. Let us first recall that
the Poisson bracket relation that defines the symplectic product is the same
for the torus as in the case of the plane. Therefore the classical generating
function for canonical transformations are governed by the same composition
laws as defined in \cite{ozrep} for the plane case. The only difference is
that there would be different chords for a given center due to the periodic
boundary conditions that identify centers with half the period as that of
the whole torus.

We will then study dynamical systems with $L$ degrees of freedom for which
there is defined a Hamiltonian function that generates the dynamics through
Hamilton's equations and that is periodic in all its $2L$ variables. For $%
L=1 $ this kind of system has applications in solid state physics; it has
been used to model electron eigenstates in a one-dimensional solid with an
incommensurate modulation of the structure \cite{aubry} and in models of
Bloch electrons in a magnetic field \cite{harp}. It has also been shown \cite
{willk} that this model presents a critical behavior giving rise to
hierarchical structures in the solutions throughout the spectrum of the kind
known as a {\it Hofstadter butterfly} \cite{hofs} and to localization
transitions from extended to localized states.

The Fourier theorem ensures that a classical Hamiltonian that is periodic in
the plane can be written as 
\begin{equation}
H(p,q)=\sum_{r,s=-\infty }^{+\infty }H_{r,s}e^{i2\pi (rp-sq)}.
\end{equation}
To quantize this Hamiltonian, different ways may be taken involving
different orderings. We choose the Weyl ordering, which is such that 
\begin{equation}
\widehat{H}(\hat{p},\hat{q})=\sum_{r,s=-\infty }^{+\infty }H_{r,s}e^{i2\pi (r%
\hat{p}-s\hat{q})}.
\end{equation}
With the definition of the translation operators on the plane (\ref{eq:tcor}%
), we immediately see that this is equivalent to 
\begin{equation}
\widehat{H}(\hat{p},\hat{q})=\sum_{r,s=-\infty }^{+\infty }H_{r,s}\widehat{T}%
_{\xi _{r,s}}  \label{eq:hqt}
\end{equation}
where $\xi _{r,s}=(2\pi \hbar r,2\pi \hbar s)=(\frac rN,\frac sN)$. The
Hamiltonian is then a linear combination of translation operators that
leaves ${\cal H}_N^\chi $ invariant. If another ordering is chosen, there
will be corrections to (\ref{eq:hqt}) of the order of $\frac 1N$.

The quantal evolution of the system is determined by the propagator:

\begin{equation}
\widehat{U}_t=e^{\frac i\hbar t\widehat{H}}=\sum_{n=0}^\infty \frac
1{n!}(\frac i\hbar t\widehat{H})^n.
\end{equation}
This last relation implies that the propagator is a combination of products
of torus translations in the expansion (\ref{eq:hqt}). These form a cocycle,
as we already saw, so we can write 
\begin{equation}
\widehat{U}_t=\sum_{r,s=-\infty }^{+\infty }U_{r,s}\widehat{T}_{\xi _{r,s}}.
\end{equation}
Thus, the evolution operator also leaves ${\cal H}_N^\chi $ invariant.
Written in this way we can see that the evolution operator and the
Hamiltonian in the periodic plane have their chord representation in terms
of torus translations only, in the form

\begin{equation}
U_t(\xi )=\sum_{r,s=-\infty }^{+\infty }U_{r,s}\delta (\xi_{r,s} -\xi).
\end{equation}
Let us now project this operator on ${\cal H}_N^\chi $ and follow the
evolution. We may first note that (\ref{eq:abpt}) and (\ref{eq:hqt}) allows
us to write:
\begin{eqnarray}
\widehat{{\bf U}}_t &=&\widehat{U}_t\widehat{{\bf 1}}_N^\chi
=\sum_{r,s=-\infty }^{+\infty }U_{r,s}\widehat{{\bf T}}_\xi
=\sum_{n=0}^\infty \frac 1{n!}(\frac i\hbar t\widehat{H})^n\widehat{{\bf 1}}%
_N^\chi  \nonumber \\
&=&\sum_{n=0}^\infty \frac 1{n!}(\frac i\hbar t\widehat{H}\widehat{{\bf 1}}%
_N^\chi )^n=e^{\frac i\hbar t{\bf \hat{H}}} , \label{eq:UHT}
\end{eqnarray}
where ${\bf \hat{H}=}\widehat{{\bf 1}}_N^\chi \hat{H}\widehat{{\bf 1}}%
_N^\chi =\hat{H}\widehat{{\bf 1}}_N^\chi $ is the Hamiltonian acting in the
torus Hilbert space ${\cal H}_N^\chi .$

The unitarity evolution operators form a group, such that 
\begin{equation}
\widehat{U}_t=(\widehat{U}_{\frac tM})^M.
\end{equation}
Projecting onto the torus and using (\ref{eq:abpt}) we obtain 
\begin{eqnarray}
\widehat{{\bf U}}_t &=&\widehat{{\bf 1}}_N^\chi \widehat{U}_t\widehat{{\bf 1}%
}_N^\chi =\widehat{{\bf 1}}_N^\chi (\widehat{U}_{\frac tM})^M\widehat{{\bf 1}%
}_N^\chi  \label{eq:prodevol1} \\
&=&(\widehat{{\bf U}}_{\frac tM})^M.  \label{eq:prodevol2}
\end{eqnarray}
This last result is very important; the evolution and the projection commute.

We have two alternatives to obtain the center representation for (\ref
{eq:prodevol2}) , i.e. to work from the plane relations or to work directly
with the torus. First we note that (\ref{eq:UHT}) implies 
\begin{equation}
\lim_{t\rightarrow 0}{\bf U}_t(x)=e^{\frac i\hbar t{\bf H}(x)}+0(t^2).
\end{equation}
The use of (\ref{eq:prodevol2}) and (\ref{3.77}) results in

\begin{eqnarray}
{\bf U}_t(x) &=&\lim_{M\rightarrow \infty }\left( \frac 1N\right)
^{2LM}\sum_{x_i=0}^{\frac{N-1}2}f_N(x+\sum_{j=1}^{2M}(-1)^jx_j)  \nonumber \\
&&\exp \left\{ \frac i\hbar \left[ \,\Delta _{2M+1}(x,x_1,...,x_{2M})-\frac
t{2M}\sum_{i=1}^{2M}{\bf H(}x_i)\right] \right\} .  \label{eq:utxT}
\end{eqnarray}
For the odd $N$ case we obtain the representation on the points $X$ of QPS, 
\begin{equation}
{\bf U}_t(X)=\lim_{M\rightarrow \infty }\left( \frac 1N\right)
^{2LM}\sum_{X_i=0}^{N-1}\exp \left\{ \frac i\hbar \left[ \,\Delta
_{2M+1}(X,X_1,...,X_{2M})-\frac t{2M}\sum_{i=1}^{2M}{\bf H(}X_i)\right]
\right\} .  \label{eq:utxTni}
\end{equation}
Notice that this expression for the projector relies on our original product rule for an arbitrary number of operators.

To take the projection, using (\ref{eq:prodevol1}), we can use the already
known result about the propagator in the center representation on the plane 
\cite{ozrep}, obtained as a {\it path integral} 
\begin{eqnarray}
U_t(x) &=&\lim_{M\rightarrow \infty }\int \frac{dx_1\cdots dx_{2M}}{(\pi
\hbar )^{2ML}}\ \exp \left\{ \frac i\hbar [\Delta _{2M+1}(x,x_1,\cdots
,x_{2M}))-\frac tM\sum_{n=1}^{2M}H(x_n)]\right\}  \label{eq:utx2} \\
&=&\int_\gamma {\cal D}_\gamma \,e^{i/\hbar S_\gamma (x)}.
\end{eqnarray}
Here we can see that for the odd $N$ case the propagator (\ref{eq:utxTni})
is similar to (\ref{eq:utx2}) replacing the integral by the appropriate
sums. The phase of the integral in (\ref{eq:utx2}) coincides with the center
action $S_\gamma (x)$ for the polygonal path $\gamma $ with endpoints
centered on $x$ and whose $k^{\prime }$th side is centered on $x_k$. The
center variational principle ensures that this center action is stationary
for the classical trajectories centered on $x$. In Fig.~\ref{fig.1.8.}. we show two
possible paths, whose actions are compared by the center variational
principle.
\begin{figure}[h]
\vspace{7cm}
\caption{ Two possible paths, whose actions are compared by the center
variational principle.
}
\label{fig.1.8.}
\end{figure}

Now we project the symbol on the torus through (\ref{eq:Axplator}). We then
obtain 
\begin{eqnarray}
{\bf U}_t(x) &=&\left( \frac 12\right) ^L\left\langle e^{i2\pi N(\left[
\left( x-\frac \chi N\right) \wedge {\bf k+}\frac 14{\bf k}\widetilde{{\frak %
J}}{\bf k}\right] )}\lim_{M\rightarrow \infty }\int \frac{dx_1\cdots dx_{2M}%
}{(\pi \hbar )^{2ML}}\right.  \nonumber \\
&&\left. \exp \left\{ \frac i\hbar [\Delta _{2M+1}(x+\frac k2,x_1,\cdots
,x_{2M}))-\frac t{2M}\sum_{n=1}^{2M}H(x_n)]\right\} \right\rangle _{{\bf k}}
\nonumber \\
&=&\left( \frac 12\right) ^L\left\langle e^{i2\pi N(\left[ \left( x-\frac
\chi N\right) \wedge {\bf k+}\frac 14{\bf k}\widetilde{{\frak J}}{\bf k}%
\right] )}\int_\gamma d\gamma \,\;e^{\frac i\hbar S_\gamma (x+\frac{{\bf k}}%
2)}\right\rangle _{{\bf k}}.  \label{eq:utxT2}
\end{eqnarray}
Although it is not immediately evident, (\ref{eq:utxT}) and (\ref{eq:utxT2})
are the same object; in (\ref{eq:utxT}), we first project on the torus and
then perform the evolution, while in (\ref{eq:utxT2}) we first evolve on the
plane and the projection on the torus is performed later. But, since the
projection (\ref{eq:prodevol2}) and evolution are commuting operations, (\ref
{eq:utxT}) and (\ref{eq:utxT2}) coincide. If we had defined the Weyl
transformation intrinsically in the torus, without projecting from the
plane, we could still derive a formula equivalent to (\ref{eq:utxT2}) with
the help of a Poisson transformation applied to (\ref{eq:utxT}).

To take the semiclassical limit, (\ref{eq:utxT2}) is the adequate
expression. Indeed, to apply this semiclassical limit we must evaluate the
integrals in (\ref{eq:utx2}) by the stationary phase approximation as in 
\cite{ozrep}, so 
\begin{equation}
U_t(x)_{SC}\sim \sum_j2^L|\det (1+{\cal M}_j)|^{-\frac 12}\ \exp \left\{
i\hbar ^{-1}S_{tj}(x)+i\mu _j\right\} ,  \label{eq:utxsc}
\end{equation}
where ${\cal M}$ is the symplectic matrix for the linearized transformation
between the neighborhood of the tips of the chord $\xi (x)$ generated by $%
S_t(x)$ as a center function.The index runs over all the contributing
classical orbits. In the case of a single orbit, the corresponding {\it %
Morse index} $\mu _j=0$. Hence on the torus we obtain 
\begin{eqnarray}
{\bf U}_t(x)_{SC} &\sim &\left\langle e^{i2\pi N(\left[ \left( x-\frac \chi
N\right) \wedge {\bf k+}\frac 14{\bf k}\widetilde{{\frak J}}{\bf k}\right]
)}\right.  \nonumber \\
&&\left. \sum_j2^L|\det (1+{\cal M}_j)|^{-\frac 12}\ \exp \left\{ i\hbar
^{-1}S_{tj}(x+\frac k2)+i\gamma _j\right\} \right\rangle _{{\bf k}}.
\label{eq:UTxsc}
\end{eqnarray}
The sum over $k$ is a sum over center points that are equivalent on the
torus because of the boundary conditions, but are different points on the
plane. To obtain the correct periodicity, the contribution of each term must
be summed with different phases. But for each point there are several
classical orbits whose chord is centered on it. The contribution of those
orbits are obtained in the $j$ sum. Then, for the semiclassical propagator
on the torus, we have a multiplicity of chords for any center point, due to
the boundary conditions.

\section{Conclusions}

\setcounter{equation}{0}

Our construction of the Weyl representation on the
torus naturally generates the conjugate chord representation.
This appears to be more useful on the torus than on the plane where it also
arises. The advantage of our derivation of the Weyl representation resides
in the clear geometrical interpretation of the operator basis in terms of
translations and reflections in QPS, so that the law for the symbol of the
product of operators acquires a simple form and generalizes to multiple products. It is important to note that
the parity of the number of states $N$ plays an important role and the
product law for $N$ odd is related to that in the plane case, by merely
replacing the integrals by the appropriate sums.

Although the geometric interpretation is valid for toral geometries, the
construction can be applied to any system whose Hilbert space has finite
dimension irrespective of the geometric structure of the underlying phase
space, except for its compactness. Indeed this operator basis and symbols
can be applied, for example, to spin systems or many-body fermionic systems 
\cite{lipgal}. However, such a generalization destroys the intuitive
interpretation of the semiclassical limit.

By defining the operators on the torus as the projections of their analogues
on the plane, some important properties of the plane can then be used on the
torus. We exploit this fact for periodic Hamiltonian systems where we map
the continuous problem on a finite dimensional one. The path integral
formulation of Hamiltonian systems on the plane allows us to obtain that on
the torus, thus illuminating the semiclassical limit.

The symplectic invariance of the Weyl representation on the plane translates
to the torus as the Feline invariance; this fact will be used to study cat
maps of general dimension \cite{loxocat}.

{\it Acknowledgments: }We thanks A. Voros, M. Saraceno and R.O. Vallejos
for helpful discussions. We acknowledge financial support from Pronex-MCT
and A.M.F.R. also thanks support from CLAF-CNPq.

\newpage


\clearpage

\begin{thebibliography}{99}
\bibitem{ozrep}  A.M. Ozorio de Almeida, Physics Report {\bf 295} (1998),
266.

\bibitem{arnold}  V.I. Arnold, "Mathematical Methods of Classical Mechanics"
(Springer, New York) (1978).

\bibitem{hanay}  J.H. Hannay and M.V. Berry, Physica {\bf 1D} (1980),
267-230.

\bibitem{balavoros}  M. L. Balazs and A. Voros, Annals of Physics {\bf 190}
(1989), 1.

\bibitem{saraceno}  M. Saraceno, Annals of Physics {\bf 199} (1990), 37.

\bibitem{pqtor}  T.S. Santhanam and A.R. Tekumalla, Found. of Phys. {\bf 6}
(1975), 5.

\bibitem{wooters}  W.K. Wooters, Annals of Physics {\bf 176 } (1987), 1.

\bibitem{galeti1}  D. Galetti and A.F.R. de Toledo Piza, Physica {\bf A 149}
(1988), 267.

\bibitem{galeti2}  D. Galetti and A.F.R. de Toledo Piza, Physica {\bf A 186}
(1992), 513.

\bibitem{kaperpeev}  W.K. Kaperskovitz and M. Peev, Annals of Physics {\bf %
230} (1994), 21.

\bibitem{debievre}  A.Bouzouina and S. De Bi\`{e}vre, Commun. Math. Phys. 
{\bf 178} (1996), 83.

\bibitem{schwinger}  J. Schwinger, Proc. Nat. Acad. Sci.{\bf \ 46 } (1960),
570 ,893,1401.

\bibitem{loxocat}  A.M.F. Rivas, A.M. Ozorio de Almeida and M. Saraceno,
''Quantization of Multidimensional Cat Maps'' Preprint.

\bibitem{keat1}  J. P. Keating, Nonlinearity {\bf 4} (1991), 277.

\bibitem{keat2}  J. P. Keating, Nonlinearity {\bf 4} (1991), 309.

\bibitem{aubry}  S. Aubry and G. Andr\'{e}, Ann. Israel Phys. Soc. {\bf 3}
(1979), 133-164.

\bibitem{harp}  P.G. Harper, Proc. Phys. Soc. {\bf A68} (1955), 874-892.

\bibitem{willk}  M. Wilkinson, Proc. R. Soc. London {\bf A 391} (1984), 305.

\bibitem{hofs}  D. R. Hofstadter, Phys. Rev. {\bf B 14} (1976), 2239-2249.

\bibitem{lipgal}  D. Galetti and A.F.R. de Toledo Piza, Physica {\bf A 214}
(1995), 207.
\end{thebibliography}
\end{document}